\documentclass[aps,prb,twocolumn,showpacs,notitlepage,superscriptaddress]{revtex4-1}
\usepackage{graphicx,amsmath,mathrsfs}
\usepackage{amssymb}
\usepackage{amsthm}
\usepackage{bm}
\usepackage{float}
\usepackage{graphicx}
\usepackage{amsmath}
\usepackage{amstext}
\usepackage{amssymb}

\usepackage{mathrsfs}
\usepackage{amsfonts}
\usepackage{amsbsy} 
\usepackage{dcolumn}
\usepackage{bm}
\usepackage{multirow}
\usepackage{color}
\usepackage{datetime}

\usepackage{hyperref}

\usepackage{epsfig}

\usepackage{color}
\usepackage{graphicx}
\usepackage{amsmath}
\usepackage{amstext}
\usepackage{amssymb}
\usepackage{amsfonts}
\usepackage{amsbsy} 
\usepackage{dcolumn}
\usepackage{bm}
\usepackage{multirow}

\usepackage{color}

\usepackage[bbgreekl]{mathbbol}
\usepackage{amscd}
\def\bc{\begin{center}}
\def\ec{\end{center}}

\newcommand{\ket}[1]{\left|#1\right\rangle}
\newcommand{\bra}[1]{\left\langle#1\right|}

\renewcommand {\ec}{\eta_{\gamma}}

\def\bra#1{\left\langle#1\right|}
\def\ket#1{\left|#1\right\rangle}

\begin{document}
\title{Free-fermion Entanglement Spectrum through Wannier Interpolation}
\author{Ching Hua Lee}
\affiliation{Department of Physics, Stanford University, Stanford, CA 94305, USA}
\author{Peng Ye}
\affiliation{Perimeter Institute for Theoretical Physics, Waterloo, Ontario, Canada N2L 2Y5}

\begin{abstract}
Quantum Entanglement plays an ubiquitous role in theoretical physics, from the characterization of novel phases of matter to understanding the efficacy of numerical algorithms. As such, there have been extensive studies on the entanglement spectrum (ES) of free-fermion systems, particularly in the relation between its spectral flow and topological charge pumping. However, far less has been studied about the \emph{spacing} between adjacent entanglement eigenenergies, which affects the truncation error in numerical computations involving Matrix Product States (MPS) or Projected Entangled-Pair States (PEPS). In this paper, we shall hence derive asymptotic bounds for the ES spacings through an interpolation argument that utilizes known results on Wannier function decay. For translation invariant systems, the Entanglement energies are shown to decay at a rate monotonically related to the complex gap between the filled and occupied bands. This interpolation also demonstrates the one-to-one correspondence between the ES and the edge states. Our results also provide asymptotic bounds for the eigenvalue distribution of certain types of Block Toeplitz matrices common in physics, even for those not arising from entanglement calculations. 
\end{abstract}
\maketitle

\tableofcontents

\section{Introduction}

Quantum entanglement has attracted intense interest in the recent years. It characterizes the amount by which classically independent bits are correlated, and has been extensively used in the study of novel phases and critical phenomena\cite{holzhey1994,PhysRevLett.96.110404,ryu2006,wolf2006,verstraete2006,plenio2005,li2008,regnault2009,laeuchli2010,yao2010,turner2010,fidkowski2010,chandran2011,qi2012}, particularly of exotic topological states\cite{kitaev2006,levin2006}. Information about the extent of entanglement between two subsystems is contained in the reduced density matrix (RDM) $\rho$, which is obtained by tracing the density matrix over one of the subsystems. The entanglement entropy (EE) is simply defined by $S=-{\rm tr}\left(\rho\log\rho\right)$, while the entanglement spectrum (ES) consists of all the eigenvalues of $\rho$, and contains more precise information on the amount of entanglement\cite{li2008}.

Very importantly, the study of Quantum Entanglement has also revolutionized the development of numerical computational methods, particularly those for gapped systems which exploit the short-range entanglement between Matrix Product States (MPS) in one dimension, and Projected Entangled-Pair states (PEPS) in higher dimensions\cite{kraus2010,dubail2013,wahl2013,wahl2014}. These methods have a long legacy of triumphs, starting from the now ubiquitous density-matrix renormalization group (DMRG) algorithm\cite{white1992} . With them, the physical properties of a large class of systems are computed with hirtherto unattained accuracy and efficiency\cite{peschel1999,schollwock2005}. In essence, it drastically reduces the complexity of the calculation by discarding irrelevant degrees of freedom (DOFs) in a Schmidt decomposition. However, there is a trade-off between gains in computational efficiency and the truncation error accured, the latter which is bounded by the decay rate of the entanglement spectrum (ES). Indeed, it is of practical importance to have an analytic understanding of the asymptotic properties of the entanglement spectrum.  

As such, the main objective of this work will be to concretely understand the asymptotic decay properties of the ES of free fermion systems. Our approach involves an explicit interpolation between the ES of the system and its Wannier polarization spectrum\cite{marzari1997, qi2011}. This interpolation gives a physically intuitive picture relating wavefunction localization and their entanglement properties,  where the imaginary gap that controls the wavefunction locality is shown to also provide a rigorous lower bound for the decay rate of the ES. It also gives a natural explanation for the edgestate-entanglement spectrum correspondence that has already garnered significant interest in the study of topological condensed matter systems\cite{qi2012,hughes2011}. While entanglement studies based on the aforesaid interpolation already exist in the literature\cite{alexandradinata2011,huang2012}, they were primarily concerned about similarities betweem the topological behavior of the Wannier functions and the ES, and not their decay properties whose quantitative study is the focus of this work.

The value of our analytic asymptotic bounds on the ES potentially extend beyond problems on entanglement. This is because mathematically, the ES of free fermions corresponds to the eigenspectra of a certain class of matrices known as Block Toeplitz matrices with singular symbols (see Sect. \ref{sec:toeplitz} for a definition) which are ubiquitous in various areas of physics, whenever there are translationally-invariant systems with internal DOFs and abrupt truncations. They appear, for instance, in various spin chain models\cite{its2005, its2008, keating2005}, dimer models\cite{basordimer2007}, impenetrable bose gas systems\cite{ffapps} and full counting statistics pertaining to certain non-equilibrium phenomena involving quantum noise \cite{ivanov2010phase,noneq}. But due to considerable mathematical difficulties, there has been no known explicit result for the asymptotic eigenspectra of such Toeplitz matrices, except for the simplest few cases\cite{its2008}. As such, we hope that our asymptotic results will shed some additional light onto the solutions of a wealth of physical problems, despite being just asymptotic bounds. The reader is invited to read Appendix \ref{app:history} for more details on the illustrious history of Toeplitz matrices.

In this paper, we shall derive estimates for the asymptotic spacing between entanglement energies for generic free-fermion lattice systems. Despite being asymptotic results, these estimates are in practice quite accurate beyond the first one or two eigenenergies. Inspired by the edge spectrum - entanglement spectrum correspondence suggested in Refs. \onlinecite{qi2012,li2008,thomale2010,yao2010,turner2010,fidkowski2010,PhysRevB.86.045117}, we constructed an explicit interpolation between the Wannier operator and the single-particle correlator. This interpolation, which is the highlight of this work, provides a physically-motivated explanation of the relation between the decay rate in the ES and that of the Wannier spectrum. While our main results do not require the system to be translationally invariant, if the latter condition holds the essential behavior of the ES can be directly expressed in terms of the complex-analytic properties of the lattice hamiltonian, the same properties that govern the spatial rate of decay of the Wannier functions.

This paper is organized as follows. In section \ref{sec:foundations}, we shall introduce the entanglement spectrum and Wannier polarization spectrum of free fermion systems, and suggest how they may be related. Following that will be Section \ref{sec:main}, where we present Eq. \ref{key}, our key result for the asymptotic ES spacing. We shall illustrate it through a toy example involving the Dirac Model, and prove it in detail via an interpolation between the Wannier operator and the entanglement projector. Finally, we shall discuss further applications of our results to the study of Block Toeplitz matrices in Section  \ref{sec:toeplitz}.

\section{Theoretical Foundations}
\label{sec:foundations}

\subsection{Entanglement Spectrum}

Consider a free-fermion system described by a Hamiltonian $H=\sum_{i,j}f_i^\dagger h_{ij}f_j$, with $f_i$ annihilating a fermion at site $i$. To study its entanglement properties, we introduce a real-space partition by defining a subregion $A$ and its complement $B=\bar{A}$ in the system. With these regions, we can define a reduced density matrix (RDM) $\rho_A$ by partially tracing out the degrees of freedom (DOFs) of region $B$:
\begin{eqnarray}
\rho_A={\rm tr_B}\left[\ket{G}\bra{G}\right]\,,
\end{eqnarray}
where $\ket{G}$ is the groundstate of the system, and $\rho=\ket{G}\bra{G}$ its full density matrix. For free fermion systems, a crucial simplification follows from the fact that all multi-point correlation functions obey Wick's theorem. This allows the following Gaussian form\cite{peschel2002} for RDM $\rho_A$:
\begin{eqnarray}
\rho_A=e^{-H_E},~H_E=\sum_{i,j\in A}f_i^\dagger {h_E}_{ij}f_j
\end{eqnarray}
where $h_E$, known as the single-particle ``entanglement Hamiltonian'', has a role superficially resembling that of a physical hamiltonian at finite temperature. This is further elaborated in Appendix \ref{peng}. Furthermore, $h_E$ can be determined from the two-point correlation function ${C}_{ij}=\bra{G}f_i^\dagger f_j\ket{G},~i,j\in A$ via
\begin{equation}
h_E=\log\left(C^{-1}-\mathbb{I}\right)
\label{peschel}
\end{equation}
with $\mathbb{I}$ the identity matrix. $C$, being the correlator \emph{within} subsystem $A$, is obtained by projecting $P$, the correlation matrix of the whole system, onto the subsystem $A$.  Writing $R=\sum_{i\in A}\ket{i}\bra{i}$ as the projection operator\cite{huang2012edge,huang2012,alexandradinata2011,lee2014prp} that implements the entanglement cut onto $A$, we obtain
\begin{equation}
\hat C= RPR\,.\notag
\label{rpr}
\end{equation}
Now, $P$ is also a projection operator, since it projects onto the occupied states via $P=\sum_n\theta(-\lambda_n)\ket{n}\bra{n}$. Here $\ket{n}$ and $\lambda_n$ are the eigenstates and eigenvalues of the single particle Hamiltonian $h$, and $\theta(x)$ is the step function. 
For instance, $P=\sum_k \theta(-\epsilon_k)$ for a Fermi sea,  
while $P=\frac{1}{2}\left(\mathbb{I}-\hat d(k)\cdot \sigma \right)$ for a two-band free-fermion lattice hamiltonian $H(k)=d(k)\cdot \sigma$, where $\sigma_i$, $i=1,2,3$ are the Pauli matrices. 

Although $P$ and $R$ do not generically commute, the eigenvalues of the operators $RPR$ and $PRP$ are in fact equal because both $P$ and $R$ are projectors. This useful little fact was shown in Refs. \onlinecite{huang2012edge,huang2012,lee2014prp}, and in fact holds for generic basis-independent combinations of $P$ and $R$. To facilitate the Entanglement-Wannier interpolation that we shall introduce shortly, we shall henceforth identify the correlator $C$ with
\begin{equation}
\hat C'=PRP
\label{prp}
\end{equation}
with the entanglement spectrum, i.e. the eigenspectrum of $h_E$, completely determined by the eigenspectrum of $\hat C$ or $\hat C'$ via Eq. \eqref{peschel} or Eq. (\ref{prp}).

\subsection{Wannier Polarization Spectrum}

We next define the Wannier polarization spectrum. The Wannier functions $\ket{\psi}$ are defined as the eigenfunctions of the \emph{Wannier operator}\cite{kivelson1982} 
\begin{equation}
\hat W= PXP
\label{wannier}
\end{equation}
where $P$ is the projectors onto the occupied bands as before, and $X=\frac{x}{L}$ is the position operator that takes values between $0$ and $1$, where $L$ is the length of the system in the direction of $x$. The eigenvalues of $\hat W$ form the \emph{Wannier polarization spectrum}, which physically correspond to the centers of mass of the corresponding Wannier functions (WFs) $\psi(x)=\langle x|\psi\rangle$, as plotted in Fig. \ref{comparespectra}. Essentially, the latter are the `best possible' localized orbitals formed from the occupied DOFs, and will reduce to delta function peaks when there are no unoccupied bands, i.e. when $P$ is trivial. That the WFs are indeed maximally localized has been shown in various sources like Refs. \onlinecite{kivelson1982,marzari1997}. For our purposes, their optimal localization allows us to uniquely determine their real-space decay rate which we shall utilize extensively later on. Note that a periodic version of $\hat W$, i.e. with $X=e^{\frac{2\pi i x}{L}}$, is often used in the literature instead\cite{qi2011,lee2013,lee2013lattice,yu2011}, in order to be consistent with the periodicity of the system. In our case, however, it is more convenient to use the aperiodic version from Eq. \ref{wannier} since we will be studying the physics near the entanglement cut.

\subsection{Comparison of $\hat C'$ and $\hat W$}

\begin{figure}[H]
\includegraphics[scale=0.22]{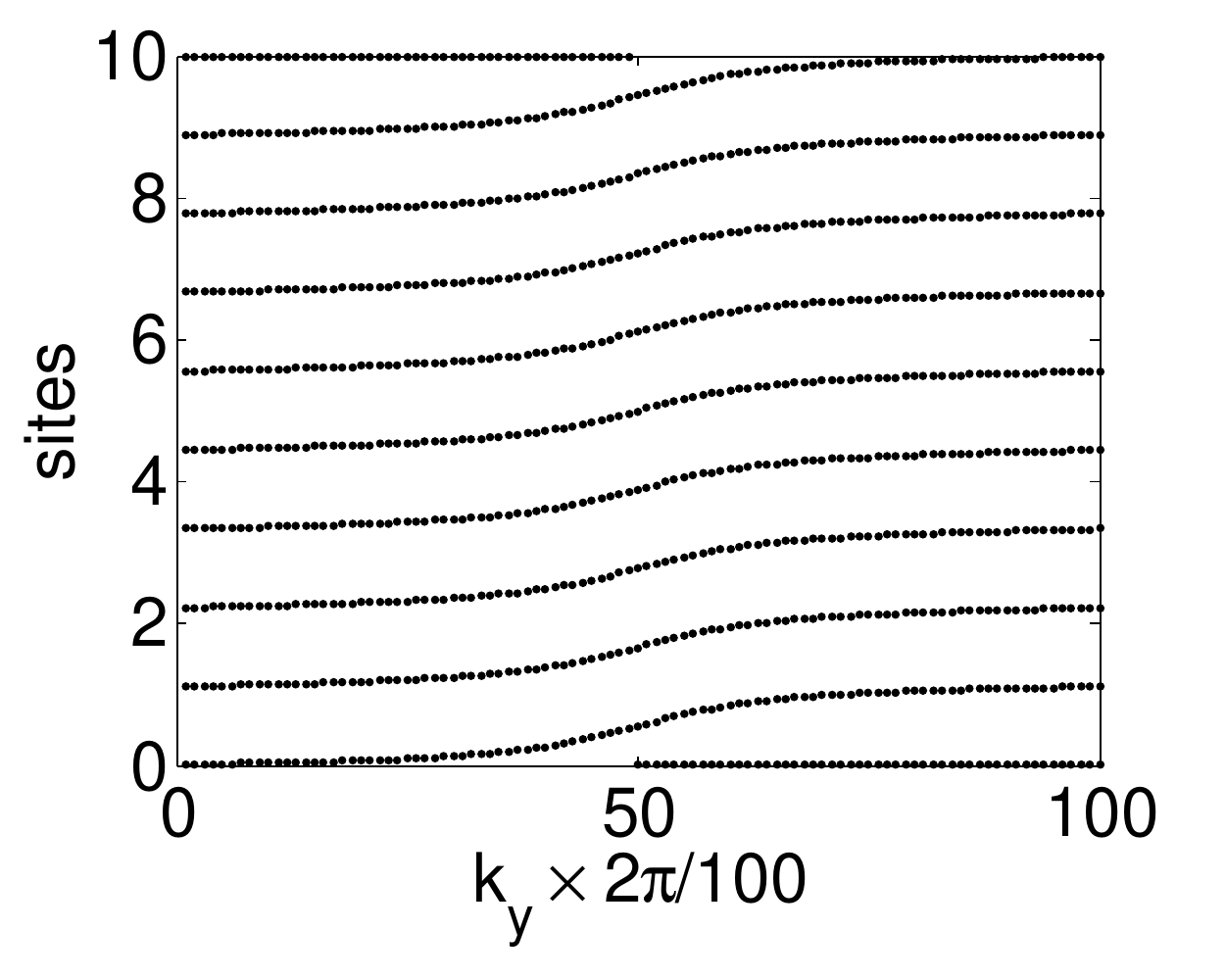}
\includegraphics[scale=0.22]{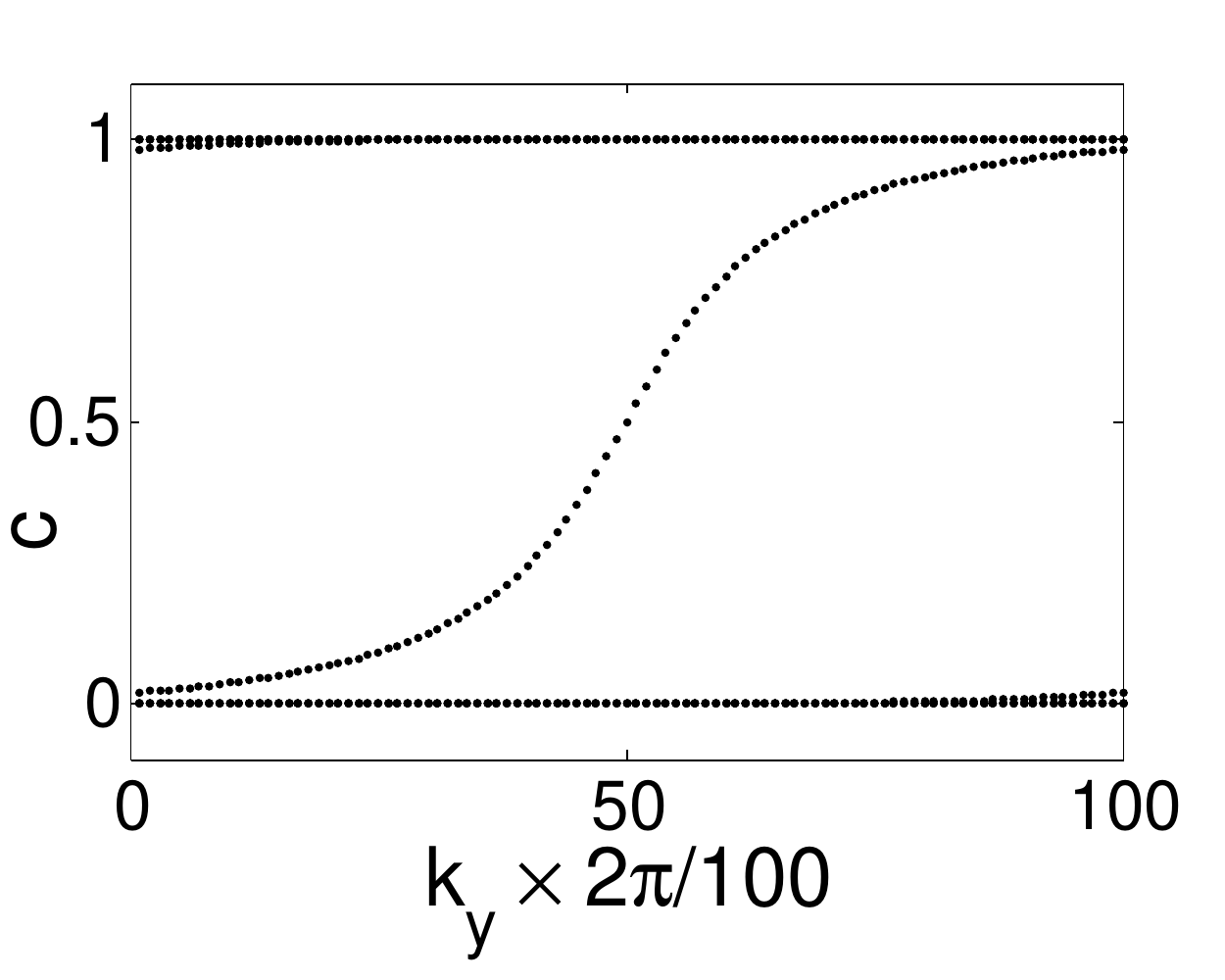}
\includegraphics[scale=0.22]{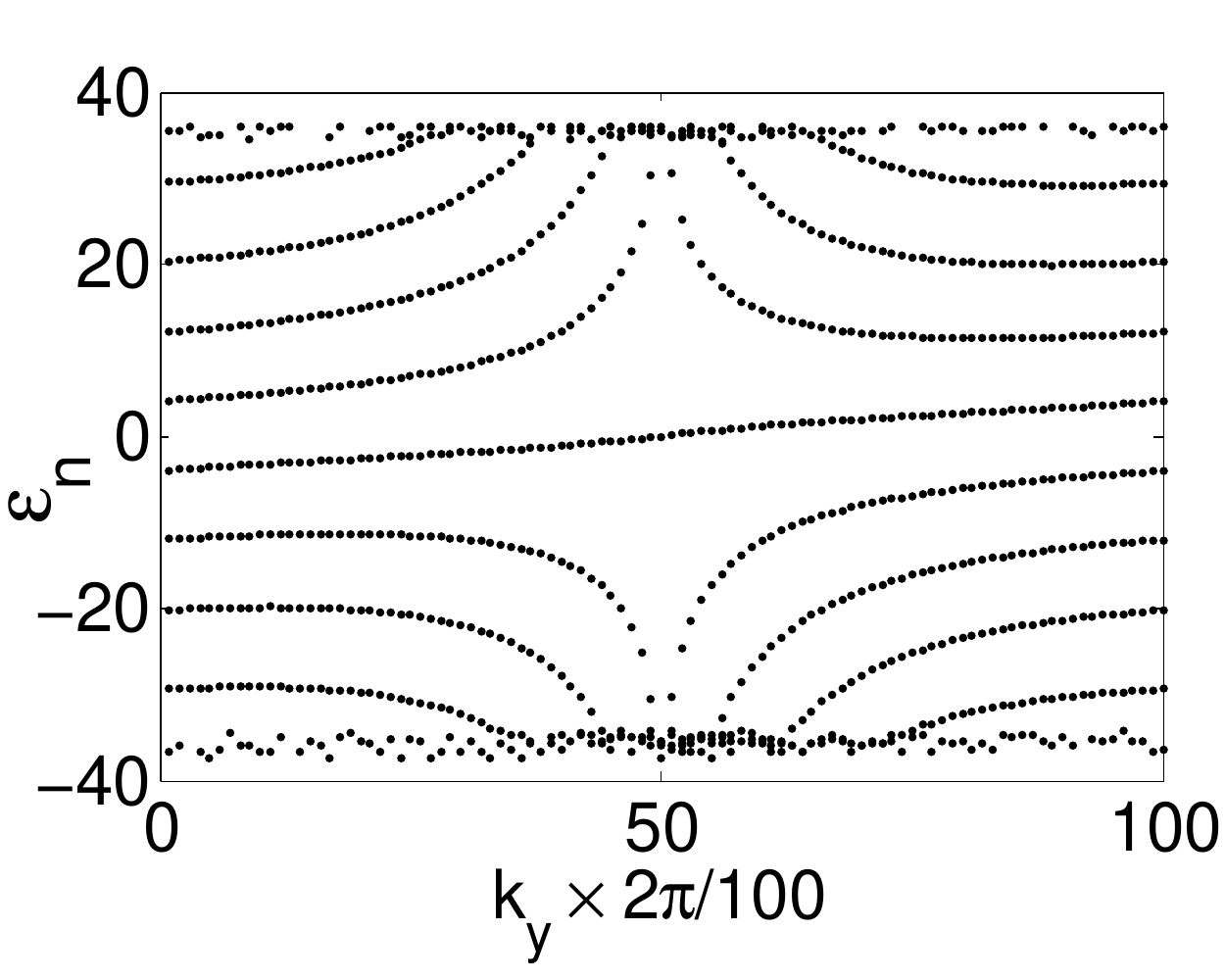}
\caption{a)  The spectrum (Wannier polarization) of $\hat W$ for the Dirac model given by Eq. \ref{dirac} with $m=1$. We see a spectral flow of $C_1=1$ site per period of $k_y$. b) The spectrum of $\hat C'$ for the same model. With the exception of one eigenvalue belonging to the edge state that exhibits an analogous spectral flow from $c=0$ to $1$, the rest stay exponentially close to $0$ and $1$, i.e are exponentially contained in one entanglement partition. c) Plot of the ES $\epsilon=\log(c^{-1}-1)$, which shows the eigenvalues $c$ very near $0$ or $1$ more clearly. The ES looks suggestively similar to the Wannier polarization, with the same spectral flow, except that the eigenvalue spacings depend on $k_y$. For clarity, we have used open boundary conditions (BCs), so that only one edge state appears. Periodic BCs will be used in subsequent plots.}
\label{comparespectra}
\end{figure}

Evidently, the entanglement correlator $\hat C'=PRP$ and the Wannier Polarization operator $\hat W=PXP$ assume similar mathematical forms although their physical interpretations are quite different. Their only difference is that $R$ is a step function in real space, while $X$ is a linear function. Their spectra are compared in Fig. \ref{comparespectra}. In the rest of this paper, we shall explore in depth the implications of interpolating between these two operators. 

\section{Main results of the Entanglement-Wannier correspondence}\label{sec:main}



Here, we consider a generic D-dimensional free-fermion system, and define the entanglement cut and the Wannier operator to be along the same direction. For now we shall assume that the system is translationally invariant before the cut, so  the crystal momenta is well-defined in the perpendicular directions, and will  collectively denoted as the $k_\perp$ parameter. The result for broken translational symmetry will be discussed at the end of this section.

\subsection{The key result}

Our \emph{key} result is that the entanglement spectrum inherits the spectral flow of the Wannier polarization spectrum, but with the gap between eigenvalues related to the imaginary gap of the system. This will be shown via the interpolation between $\hat C'$ and $\hat W$ in Section \ref{interpolation}. Quantitatively, we write
\begin{equation}
\epsilon_{n,a}(k_\perp)\approx [n+X_a(k_\perp)]f(g(k_\perp)), ~f(g)>2g
\label{key}
\end{equation}
where $\epsilon_{n,a}$ is the $n^{th}$ entanglement eigenenergy corresponding to the band/edge $a$, and $X_a(k_\perp)$ is its Wannier polarization (center-of-mass). $f(g)$ is a monotonically increasing function bounded below by $2g$, where $g(k_\perp)$ is the decay rate of the Wannier functions (WFs) that can be rigorously computed. 

Let us first briefly comment on the salient features of result Eq. \ref{key}. It states that the ES is approximately equally spaced, as shown in Fig. \ref{fig:dirac}, with the spacing depending monotonically only on the Wannier decay rate $g$. Physically, $g$ characterizes the maximal possible localization of the wavefunction using the available (occupied) states. Since entanglement measures the corresponding quantum uncertainty behind a real-space cut, it should depend monotonically with the amount of the wavefunction 'leaking' through the cut, which is quantified by $g$.   

The WS inherits a spectral flow from the Wannier polarization $X_a$. This flow arises inevitably due to a topological charge pumping mechanism, and has already been thoroughly studied in other works\cite{yu2011,huang2012edge,huang2012,alexandradinata2011}. In the following Section \ref{interpolation}, we shall justify this inheritance of spectral flow through an interpolation argument between $\hat C'$ and $\hat W$. Our simple interpolation argument provides yet another `proof' of the edge state - entanglement spectrum correspondence explored in some other works mentioned in the introduction, together with important quantitative estimates of the decay properties of the ES.

\subsection{Example: 2-D Dirac Model}\label{sec:model}

To make the above statements more concrete, we shall study the example of a 2D Dirac model with band Hamiltonian

\begin{equation}
H_{Dirac}(k)=d(k)\cdot\sigma
\label{dirac}
\end{equation}
where $\sigma_i$, $i=1,2,3$ are the Pauli Matrices, and $d(k_x,k_y)=(m+\cos k_x+\cos k_y,\sin k_x,\sin k_y)$. This is among the simplest model that exhibits a nontrivial 1-parameter spectral flow due to nontrivial topology when $|m|<2$. WLOG, we shall assume that the cut be normal to the x-direction, so that $k_\perp=k_y$ is a good quantum number.
\begin{figure}[H]
\includegraphics[scale=.49]{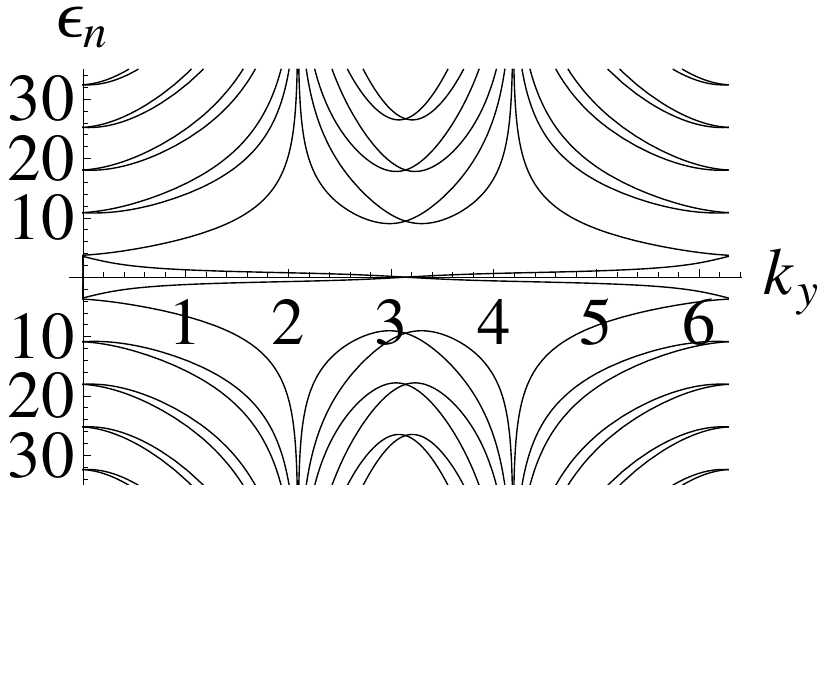}
\includegraphics[scale=.43]{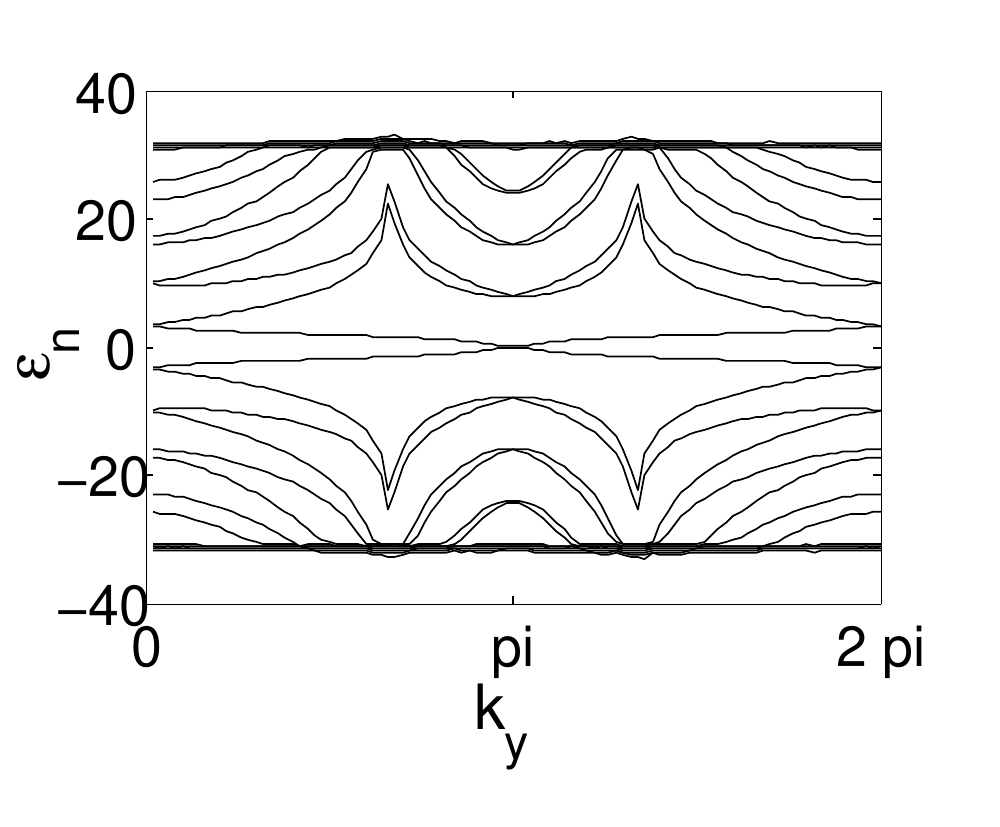}
\caption{Analytical (Eq. \ref{key}) (left) and numerical (right) results for the entanglement spectrum $\epsilon_n$ for the Dirac model with $m=0.5$. The x-axis represents $ky\in[0,2\pi]$ while the y-axis represents the entanglement energies.
Only the first few eigenvalues are plotted. The spectra agree qualitatively, and in fact exactly at $k_y=0$ and $\pi$. In the numerical plot, $\epsilon_n$ is computed down to the limits of machine precision at $\approx \pm 30$. $k_y$ is discretized into $100$ points for the numerical plot.
}
\label{fig:dirac}
\end{figure}

In Fig. \ref{fig:dirac}, the analytic approximation to the entanglement eigenvalues $\epsilon_n(k_y)$ from Eq. \ref{key} is compared against exact numerical results. We see that they agree rather well, especially for those further from zero. This is encouraging, because the decay rate $g$ in Eq. \ref{key} is exact in the asymptotic limit of large $n$, which is numerically inaccessible.

To first order, $f(g)$ may be (rather accurately) represented by a simple linear ansatz
\begin{equation}
f(g(k_y))= (2+A)g(k_y)+J
\label{linearansatz}
\end{equation}
where $A$ and $J$ are parameters that can be exactly determined by exactly evaluating ES at $k_y=0$ or $\pi$. These are two points where there exists exact analytic results for the Block Toeplitz Matrices corresponding to the ES\cite{its2005,its2009}. This will be derived in detail in Appendix \ref{2bandtoeplitz}. The physical interpretations of $A$ and $J$ will be discussed in the context of the interpolation argument in Section \ref{interpolation}.

We also observe spectra flow of the entanglement eigenvalues in Fig. \ref{fig:dirac}, which is present due to the nontrivial topology of $H_{Dirac}$. The two sets of ES eigenvalues, one advancing by one site and one receding by one site, represent the ES spectral flow of the two entanglement cuts. Similar observations are also discussed at length in Refs. \onlinecite{qi2008,hughes2011}. Like the Wannier polarization, the ES shifts by $C_1$ sites upon one periodic evolution of $k_y$, where $C_1$ is the Chern number ($C_1=1$ here) of the hamiltonian\cite{yu2011,huang2012edge,huang2012,alexandradinata2011}. As suggested by the $X_a(k_\perp)$ term in Eq. \ref{key}, this spectra flow is inherited from the Wannier polarization, which we shall quantitatively derive in Section \ref{app:dirac}.

\subsection{The Entanglement-Wannier Interpolation}
\label{interpolation}

In this subsection, we shall present the interpolation between $\hat C'$ and $\hat W$ in detail, and justify our main result Eq. \ref{key}.

\subsubsection{Definition of the interpolation}

The interpolation operator (which is slightly different from that in Ref. \onlinecite{huang2012}) is given by 
\begin{equation}
\hat W(s)=PX(s)P=P[\tilde AX_A\tilde A+\tilde B X_B\tilde B]P\,,
\end{equation}
where $\tilde A$ and $\tilde B$ are respectively the projectors onto regions $A$ and $B$. $X_A$ and $X_B$ are their position operators given by 
\begin{equation} X_A(s)= \frac{s}{L}x \label{XA}\end{equation}
\begin{equation} X_B(s)= 1-\frac{s}{L}(L-x) \label{XB}\end{equation}

Here $\tilde A X_A\tilde A$ acts on region $A$ which includes sites $x=1,\dots,L_A$ while $\tilde BX_B\tilde B$ acts on region $B$ which includes sites $x=L_A+1,\dots,L_x$. $ X(s=1)=X$ is just the usual equally-spaced position operator linearly assigning values between $x=0$ and $1$, while $X(s=0)=R$ is the coarse-grained position operator taking only values of $0$ and $1$ in regions $A$ and $B$ respectively. Hence $\hat W(1)$ is the Wannier operator while $\hat W(0)$ is the correlator corresponding to the Entanglement Hamiltonian.


\subsubsection{Evolution under the interpolation}

Now, we study how exactly the Wannier functions morph into the eigenstates of the Entanglement Hamiltonian, so as to understand the relation between Wannier polarization and the entanglement spectrum.

First, we note an important property of the maximally localized WFs, which is that they decay exponentially, i.e. $\psi(x)\sim e^{-g|x|}$ asymptotically. Their decay rate $g$ is related to the imaginary bandgap between the occupied and empty bands, which will be elaborated later.
For most realistic hamiltonians, $g\sim O(1)$, so their WFs have very small exponential tails within a few sites of their peaks. This implies that most of the WFs, except for those straddling the cut, will be mostly contained in one region, with an exponentially small tail in the other.

As such, we can gain some insight by analyzing the contributions of the $\psi(s)$ from each region separately, where $\psi(s)$ is the eigenstate of the operator $\hat W(s)$:
\begin{equation}
\psi(s)=\psi_A(s)\oplus \psi_B(s)
\end{equation}
where $\psi_A(s)$ and $\psi_B(s)$ are nonzero only in region A and B respectively. When
$s=1$, $\psi(s=1)$ is just the Wannier function. 

Let us explore what happens when $s$ is interpolated from $1$ to $0$. For definiteness, suppose that $\psi(1)$ is mostly contained in region A, i.e $\psi_A(1)$ differs by an exponentially small extent from an eigenstate of $PX_AP$. Then 
\begin{eqnarray} 
\hat W(s) \psi &= &PX_A(s)P\psi_A\oplus PX_B(s)P\psi_B\notag\\
&\approx&  x_A(s)\psi_A\oplus PX_B(s)P\psi_B\notag\\
&=& x_A(s)PX_B(s)P\psi
\end{eqnarray}
As we tune $s\rightarrow 0$, $\psi_A(s)$ will be modified to an exponentially small extent. 
This is because the operator $X_A(s)$ remains linear in $x$, and variations of $s$ merely correspond to a rescaling of coordinates\footnote{Note that this will not be true for the edge states which straddle both regions and are not approximate eigenstates of $X_A$ or $X_B$ alone}. A rescaling just introduces a scalar multiplier, and does not change the eigenstates. 


At the end of the interpolation $s=0$, $\hat W(0)$ is just the projector onto the occupied states in region $B$:
\begin{eqnarray}
\langle \hat C'\rangle &=&\langle \psi(0) |\hat W(0)|\psi(0)\rangle \notag\\
&=& \langle \psi_A(0) |\bar A X_A(0) \bar A|\psi_A(0) \rangle +  \langle \psi_B(0) |\bar B X_B(0) \bar B|\psi_B(0) \rangle \notag\\
&=&0+\langle \psi_B(0) |\bar B X_B(0) \bar B|\psi_B(0) \rangle \notag\\
&=& \langle \psi_B (0)| \psi_B(0)\rangle
\end{eqnarray} 



While we do not yet understand how $\psi_B(s)$ evolves with the interpolation, we know that it should be approximately proportional to its value $\psi_B(1)$ at the start of the interpolation, which can be rigorously computed. Since $\psi(s=1)\sim e^{-g|x|}$ where $x$ is the displacement from its center of mass (COM), $\langle \psi_B (1)| \psi_B(1)\rangle=\int_B dx |\psi(s=1)|^2\sim e^{-2gn}$, where $n$ is number of sites the COM of $\psi$ is from the entanglement cut. The error from approximating $\psi_B(0)$ by $\psi_B(1)$ also scales like (a small power of) $e^{-gn}$. 
Hence
\begin{eqnarray}
\langle \hat C'\rangle &\approx & \langle \psi_B (0)| \psi_B(0)\rangle\notag\\
&\sim & e^{-f(g)n}
\end{eqnarray} 
where $f(g)>2g$ takes into account \emph{both} the decay rate of $2g$ from $\psi_B(1)$ before the interpolation, and an additional error introduced by the interpolation. 

\begin{figure}[H]
\includegraphics[scale=0.3]{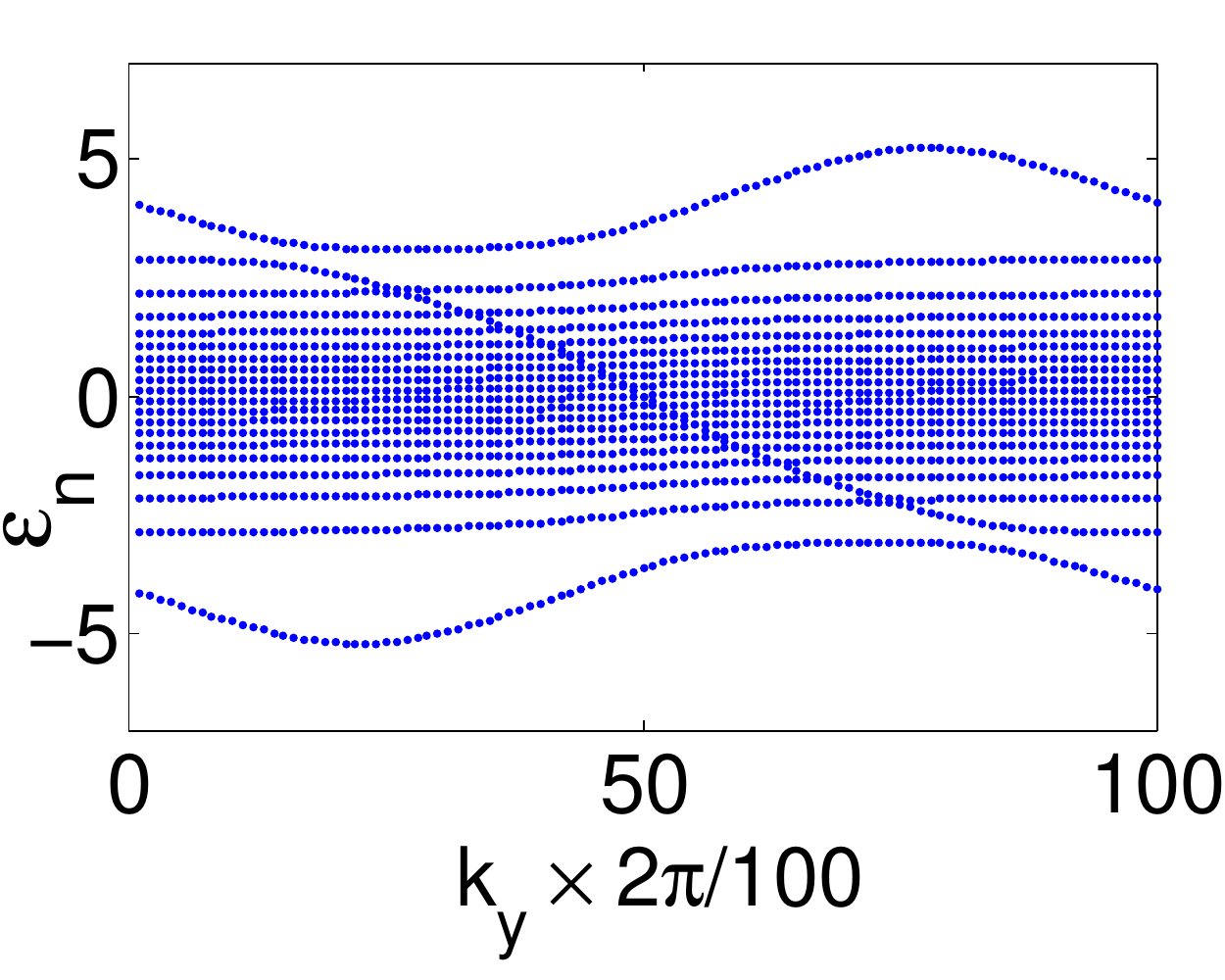}
\includegraphics[scale=0.3]{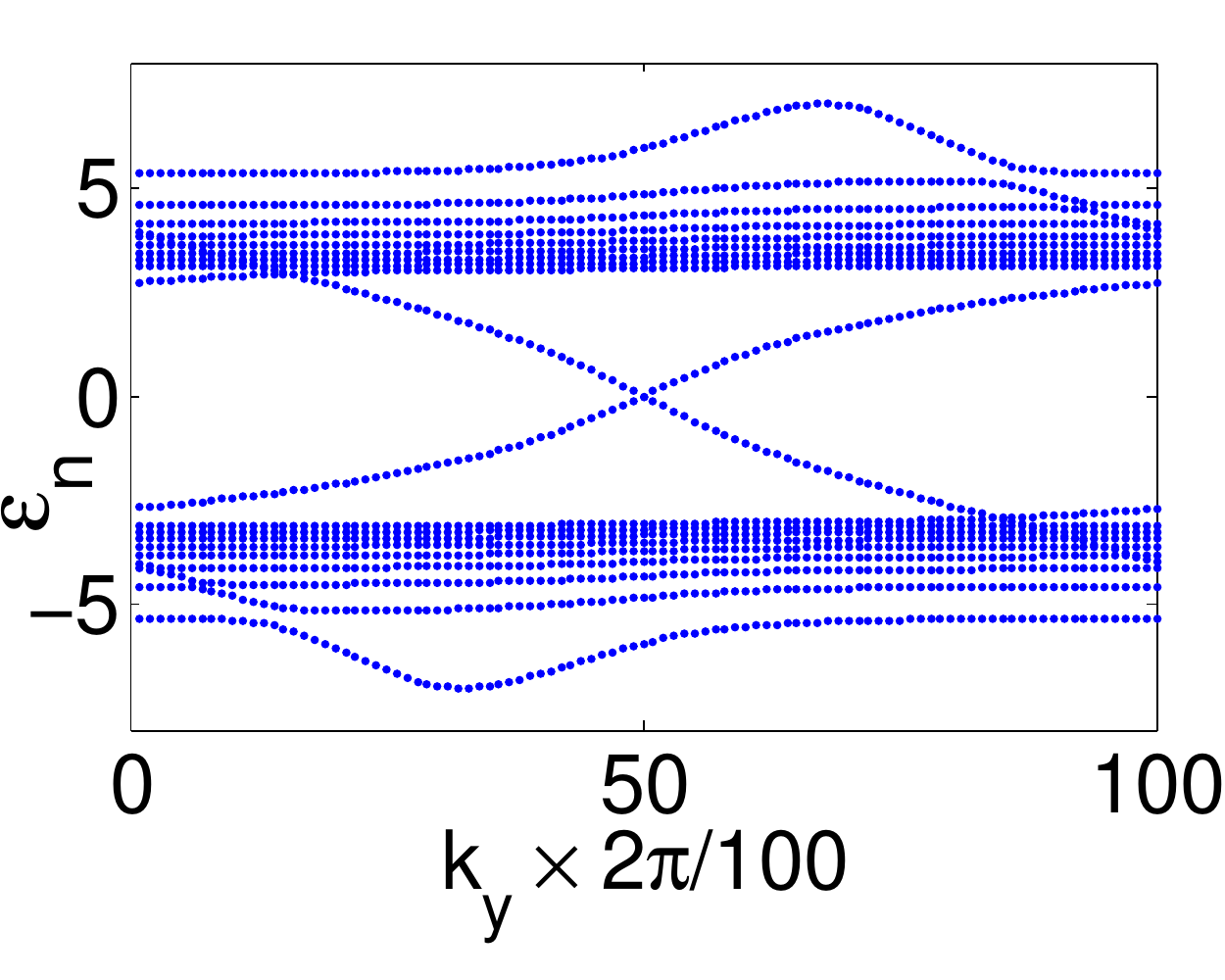}
\includegraphics[scale=0.35]{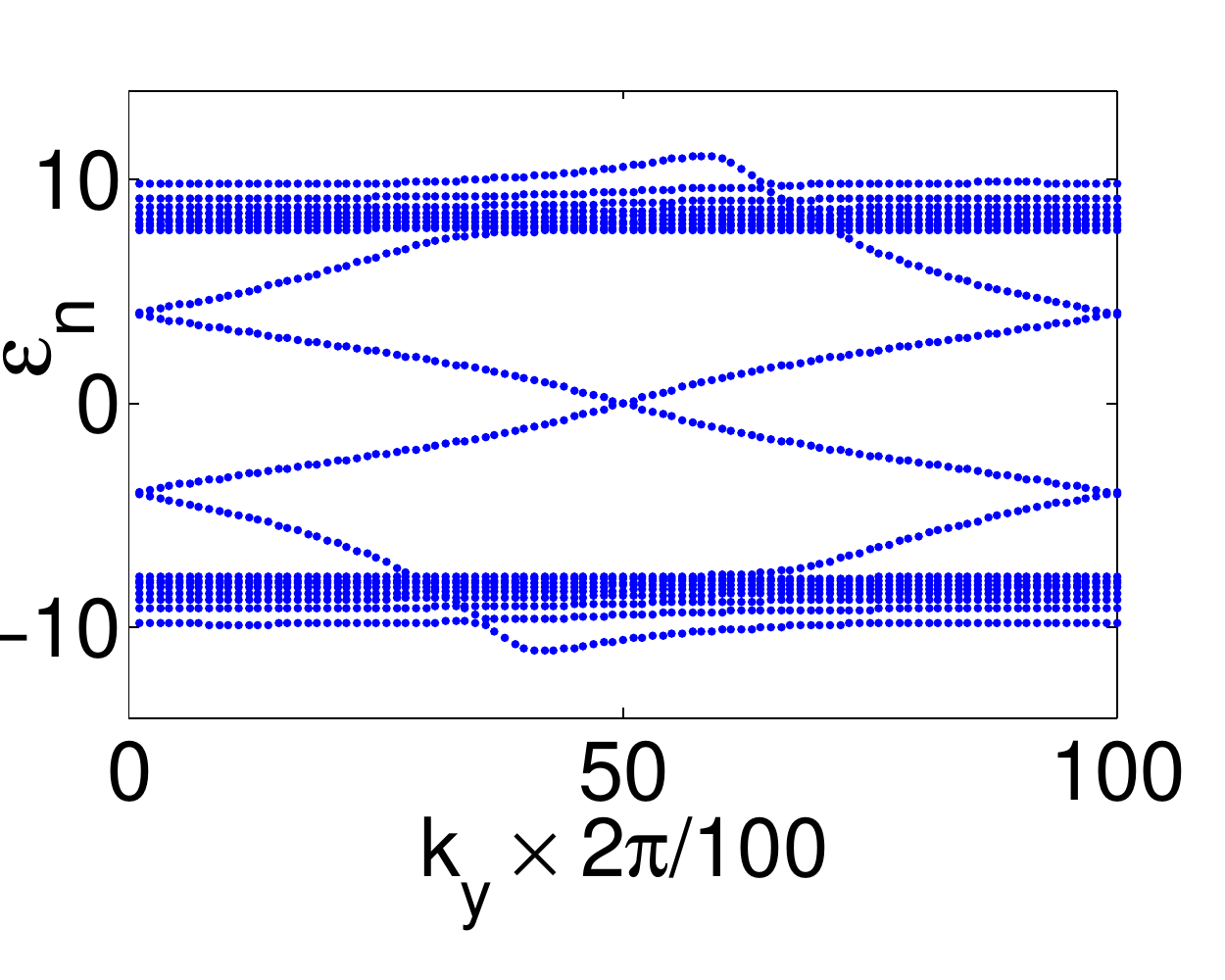}
\includegraphics[scale=0.32]{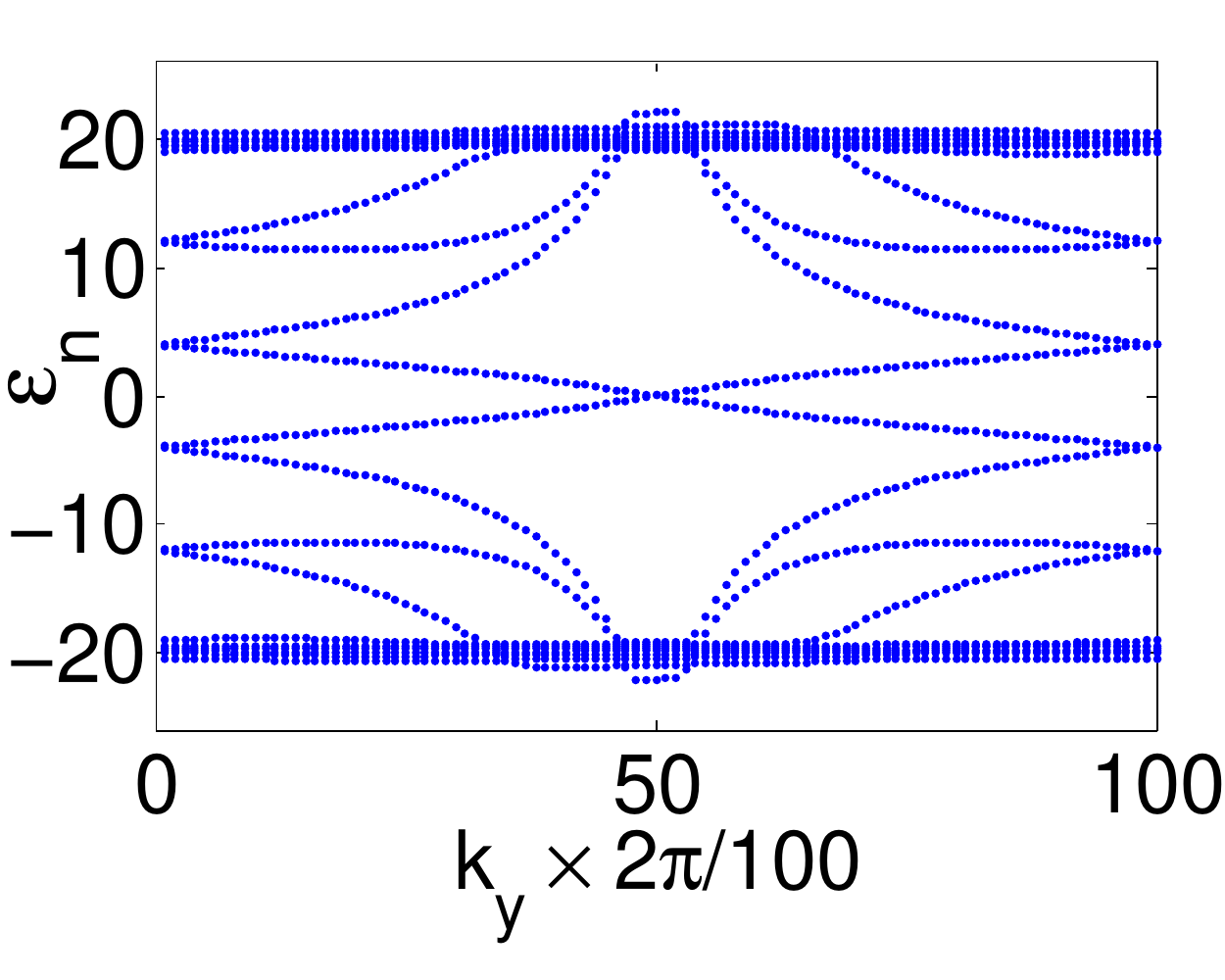}
\caption{(Top left to bottom right) Plots of $\log(w_s^{-1}-1)$, where $w_s$ are the eigenvalues of $\hat W(s)$ for $s=1,10^{-1},10^{-3}$ and $10^{-8}$. We see how the Wannier spectrum ($s=1$) evolves into the ES ($s=0$). As $s$ is decreased, $\hat W(s)$ tends towards a step function, and $w_s$ tends towards $0$ or $1$ at a rate dependent on $g(k_y)$, the rate of decay of the WFs. For $s>0$, the $w_s$'s are only exponentially spaced  exponentially spaced beneath a certain length scale set by the finite gradient of $\hat W(s)$. }
\label{fig:interpolation}
\end{figure}

Since the above interpolation is never singular, we expect a one-to-one correspondence between the Wannier spectrum and the Entanglement spectrum. Since an WF exists above each site, away from the cut the entanglement energies are, from Eq. \ref{peschel},
\begin{eqnarray}
\epsilon_n &\sim & \log(e^{f(g)n}-1)\notag\\
&\sim & f(g)n
\label{corrr}
\end{eqnarray} 
Although this linear dependence on $n$ strictly holds only for asymptotically large $n$, it holds true to better than $99\%$ for $n>2$, as evident in numerical computations (Figs. \ref{comparespectra} and \ref{fig:interpolation}). Analogous results hold when $\psi$ were mostly localized in region $B$ instead. 

In the above, it was assumed that each WF $\psi(1)$ were exactly localized $n$ sites away from the cut. In general, this may be not true, especially for topologically nontrivial systems\cite{soluyanov2011,yu2011}. We then have to replace $n$ by $n+X_a$, where $X_a$ is the Wannier polarization (shift of COM) of band $a$, yielding Eq. \ref{key}:
\begin{equation}
\epsilon_{n,a}(k_\perp)\approx [n+X_a(k_\perp)]f(g(k_\perp))
\end{equation}
where $k_\perp$ contains the momentum components transverse to the normal of the cut. 

\subsection{The Wannier decay rate $g$ elaborated}

The decay rate $g$ and hence the lower bound for the spacing of the ES can be determined precisely. A result in Fourier analysis, which will be proved in detail in Appendix \ref{fourieranalytic}, states that if
\begin{equation}
\psi(x)=\int dk e^{ikx}e^{i\theta(k)}\psi(k) 
\label{fourierresult}
\end{equation}
then $\psi(x)$ decays like $\psi(x)\sim e^{-gx}$, where $e^{i\theta(k)}\psi(k)$ has a singularity at $Im(k)=g$, but is analytic for $Im(k)<g$. When $\theta(k)$ is chosen such that $\psi(x)$ is maximally localized, both $\theta(k)$ and $\psi(k)$ depend explicitly\cite{qi2011,lee2013} only on the projector to occupied bands $P(k)$. Thus $g$ is just the distance from the real k-axis where $P(k)$ ceases to be analytic, i.e. when the gap between the occupied and unoccupied bands closes. Intuitively, a complex momentum entails a real-space decaying wavefunction because $|\psi(k)|\sim |e^{i(Re(k)+i(Im(k)))x}|\sim e^{-Im(k)x}$.

When there are only two bands, $H(k)=\sigma\cdot d(k)$ and $P(k)=\frac{1}{2}(\mathbb{I}-\hat d\cdot \sigma)$. So $g$ is simply $g=min(|Im(k_0)|)$ where $|d(k_0)|=0$. This is explicitly worked out for the Dirac Model in the Appendix. When there are more than two bands, the projector may not always be expressible in closed-form\cite{leethomale2014}. However, $g$ will always be a well-defined quantity that can be obtained numerically.

In more than one dimension, a different decay rate $g(k_\perp)$ can exist for each dimension, with $k_\perp$ denoting the momenta from the other directions. 

\subsection{Generalization to systems without translational symmetry}

We have previously focused only on translationally invariant systems with a well-defined Wannier decay rate. However, the gist of the Wannier Interpolation argument still holds true without requiring translation invariance at all. Wannier functions have well-defined decay rates even in the absence of translation symmetry, i.e. in a magnetic field where it is broken down to the magnetic translation subgroup. 

In this general setting, Eq. \ref{key} is modified to
\begin{equation}
\epsilon_{n,a_\perp}\approx \tilde f(n+X_{a_\perp}) 
\end{equation}
where $\tilde f$ is a generically nonlinear function. Here, $a_\perp$ refers to the residual collection of good quantum numbers, which can still include a transverse momentum $k_\perp$ if translation symmetry is not broken in that direction. From Eq. \ref{corrr}, the form of $\tilde f$ depends precisely on the decay behavior of the Wannier function at each position $n$ away from the cut. For instance, the orbitals of a system in a magnetic field possess a Gaussian profile, so $\tilde f$ should be a quadratic function.

\section{Relation to Eigenspectra of Block Toeplitz Matrices}
\label{sec:toeplitz}
In this short section, we shall discuss how our key result Eq. \ref{key} provides insights on the nature of the eigenspectra of certain kinds of Block Toeplitz Matrices, even those who did not originally occur in entanglement calculations. As foreshadowed in the introduction, such Toeplitz matrices are ubiquitious in diverse areas of physics. Unfortunately, exact analytic characterization of their eigenspectra is fraught with mathematical difficulties. 

Block Toeplitz matrices are finite matrices $T$ with translational invariance along each diagonal, i.e. $T_{ij}=T_{i-j}$, where each $T_{ij}$ is also a matrix which represents the internal DOFs belonging to each site. A Toeplitz matrix can be characterized its symbol, which is defined as the fourier transform along one of its rows (or columns): 
\begin{equation} g(k)=\frac{1}{2\pi}\int_{-\pi}^{\pi}T_x e^{ikx}dx \end{equation}
Loosely speaking, $g(k)$ is the 'momentum-space' representation of the matrix $T$, and a singular $g$ contains a momentum-space branch point which can be interpreted as a momentum-space projector. 


To illustrate how Toeplitz matrices appear in the calculation of entanglement spectra, we review a class of hamiltonians whose ES have been analytically studied\cite{its2009}. Consider a hamiltonian given by $H=\sigma\cdot d$, so that the projector to the occupied band $P$ is given by $P=\frac{1}{2}(\mathbb{I}-\hat d\cdot\sigma)$. If only $d_1$ and $d_2$ are nonzero, the eigenvalues $\hat \epsilon$ of $P$
can be expressed as the roots of $Det(i\hat\epsilon I + \frac{I-\hat \Gamma'}{2})$, where
\begin{eqnarray} 
\hat \Gamma'=\left( \begin{array}{cc} 
0& \frac{d_1-id_2}{\sqrt{d_1^2+d_2^2}}  \\ 
\frac{-d_1-id_2 }{\sqrt{d_1^2+d_2^2}}& 0 \end{array}\right)=\left( \begin{array}{cc} 
0& \sqrt{\frac{d_1-id_2}{d_1+id_2}}  \\ 
-\sqrt{\frac{d_1+id_2}{d_1-id_2}}& 0 \end{array} \right)\notag\\
\end{eqnarray}
For the purpose of calculating the entanglement spectrum with a cut parallel to the $y$-direction, we consign $k_\perp=k_y$ to an external parameter and consider the analytic properties of $k_x$. As $k_x$ is periodic, $d_1\pm i d_2$ will be a function of $e^{\mp ik_x}$. We can analytically continue $\hat \Gamma'$ to complex values of $k_x$ by letting it be a meromorphic function of $z=e^{ik_x}$. Since the analytic continuation is unique, $e^{-ik}\rightarrow 1/z$ not just on the unit circle where $k_x$ is real, but over the whole complex plane. All in all, we can write $\hat \Gamma'$ in terms of a degree $n$ polynomial $p(n)$ via 
\begin{eqnarray} 
\hat \Gamma'=\left( \begin{array}{cc} 
0& \sqrt{\frac{p(z)}{z^{2n}p(1/z)}} \\ 
-\sqrt{\frac{p(1/z)}{z^{-2n}p(z)}}& 0 \end{array}\right)=\left( \begin{array}{cc} 
0& \phi(z)  \\ 
-\frac{1}{\phi(z)}& 0 \end{array} \right)\notag\\
\label{toeplitzpoly}
\end{eqnarray}  
This is exactly the matrix in eq. 43 of Ref. \onlinecite{its2009}. If we want to find the entanglement entropy, we will need to project the $\hat \Gamma'$ onto region A, i.e. find the eigenvalues of $R\Gamma'R$. This can be done by fourier transforming $\hat \Gamma'$ onto real space and taking open boundary conditions. Mathematically, the real space $\hat \Gamma'(x,x')$ is a finite $2L_A \times 2L_A$ block Toeplitz matrix generated by the symbol $\hat \Gamma'(z)$ i.e.
\begin{equation}\hat \Gamma'_{ij}(x,x')=\oint \frac{\hat \Gamma'_{ij}(z)dz}{z^{x-x'+1}} \end{equation}
where $2$ is the dimension of the internal degrees of freedom and $L_A$ is the number of sites in region $A$. 

Clearly, the ES of generic hamiltonians with $N$ bands must be given by the eigenvalues of analogous Block Toeplitz Matrices with $N\times N$ blocks. The finite size of the Toeplitz matrix mathematically implements the entanglement cuts\footnote{Note that a large but finite Toeplitz Matrix will still have a qualitatively different spectrum as a truly infinite matrix.}. 

While the eigenspectrum of Toeplitz matrices without internal DOFs can be obtained rather easily through methods like Wiener-Hopf factorization
, those of \emph{Block} Toeplitz matrices (i.e. with internal DOFs) are much more elusive. In Appendix \ref{2bandtoeplitz}, we introduce some known results regarding $d_1,d_2$ of certain functional form, and then show how they, with the help of Eqs. \ref{key} and \ref{linearansatzapp}, can be directly extrapolated to more general cases involving $d_1,d_2$ and $d_3$. Indeed, the physical intuition that led to these two equations have provided a way to understand the eigenspectrum of the corresponding more general block Toeplitz matrices, whose rigorous mathematical characterization is challenging.

\section{Conclusion}

We have derived asymptotic bounds on the behavior of the Entanglement Spectrum of free-fermion lattice systems, and showed how it is related to the Wannier function decay rate which is in princple exactly computable. The asymptotic bound can be made precise when exact results are known at certain points in parameter (transverse momentum) space, as demonstated in our example. Although we have only explicitly worked out the case with two bands, the general case follows directly, with the eigenenergies from each occupied band having its own spectral flow. Similarly, our results can be extended to higher dimensions by treating the momentum in each additional transverse direction as a parameter. 
 
\begin{acknowledgements}
We thank Xiao-Liang Qi and Shuo Yang for helpful insights. C.H.L. is supported by a scholarship from the Agency of Science, Technology and Research of Singapore. P.Y. is supported by the Government of Canada through Industry Canada and by the Province of Ontario through the Ministry of Economic Development \& Innovation. P.Y. was also   supported in part by Hwa-Tung Nieh, Zheng-Yu Weng, and Tsinghua Education Foundation (North America) Inc. during his visit to KITP at Santa Barbara where the work was initiated.
\end{acknowledgements}

\appendix

\section{Relation between Entanglement Spectrum and single-particle correlation function}
\label{peng}
The relation shown in Eq. (\ref{peschel}) is a known result that first appeared in Ref. \onlinecite{peschel2002}. In the Appendix, we present its derivation in detail. 
  
 The equal-time one-particle correlation function of a fermionic system is defined as:
$ C_{i,a;j,b}=\langle f^\dagger_{i,a}f_{j,b}\rangle
$, where, $i,j,\cdots$ denote spatial coordinates, and, $a,b,\cdots$ denote spinor (for true relativistic spins) / band indices (for pseudo-spins). The average $\langle...\rangle$ is taken with respect to the ground state $|\text{GS}\rangle$ which is a \emph{pure state}.  The 2nd-quantized Hamiltonian $\mathcal{H}$ can be expressed as 
$  \mathcal{H}=\sum_{i,j,a,b}f^\dagger_{i,a} {h}_{i,a;j,b}f_{j,b}
$, where $ {h}_{i,a;j,b}$ is the single-particle Hamiltonian and fermionic operators $f,f^\dagger$ obey anticommutation relations.
In this fermionic basis, the correlation function is a hermitian matrix $  {C}$ with row indices $(i,a)$ and column indices $(j,b)$.
Let's first check the following very useful property:
\begin{align}
  C^2=  C
\end{align}
 Consider the equal-time four-point correlation function (here, $i,j,k$ temporarily incorporate internal indices for convenience):
\begin{align}
  \sum_k\langle   f^\dagger_{i} f_{k} f^\dagger_{k} f_{j}\rangle=
  \sum_k\langle f^\dagger_{i} f_{k}\rangle\langle f^\dagger_{k} f_{j} \rangle+\langle f^\dagger_i f_j\rangle\,\sum_k\langle f_k f^\dagger_k\rangle
\end{align}
by means of Wick contraction in Slater determinant wavefunction. Since the total particle number operator $\hat{N}=\sum_{i,a} f^\dagger f$ commutes with $ {H}$, indicating the $|\text{GS}\rangle$ has definite total particle number (different quantum number = different kind of particles), say $N$. Consequently, $\sum_k\langle f_k f^\dagger_k\rangle=N_0-N$ where $N_0$ is the maximum of single-particle quantum states. Therefore,
\begin{align}
  \sum_k\langle f^\dagger_{i} f_{k} f^\dagger_{k} f_{j}\rangle=
  \sum_k\langle f^\dagger_{i} f_{k}\rangle\langle f^\dagger_{k} f_{j} \rangle+\langle f^\dagger_i f_j\rangle\,(N_0-N)
\end{align}
On the other hand,
\begin{align}
  &\sum_k\langle f^\dagger_{i} f_{k} f^\dagger_{k} f_{j}\rangle=
  \langle f^\dagger_{i}(\sum_k f_{k} f^\dagger_{k}) f_{j}\rangle= \langle f^\dagger_{i}(N_0-\hat{N}) f_{j}\rangle\nonumber\\
  =&N_0\langle f^\dagger_{i} f_{j}\rangle-\langle f^\dagger_{i}\hat{N} f_{j}\rangle=N_0\langle f^\dagger_{i} f_{j}\rangle-(N-1)\langle f^\dagger_{i} f_{j}\rangle\nonumber\\
  =&\langle f^\dagger_{i} f_{j}\rangle+\langle f^\dagger_i f_j\rangle\,(N_0-N)
\end{align}
Therefore, $\sum_k\langle f^\dagger_{i} f_{k}\rangle\langle f^\dagger_{k} f_{j} \rangle=\langle f^\dagger_{i} f_{j}\rangle$, indicating that $C^2=C$.

Next, let's consider a $2+1$-D free fermion system. In $x$-$y$ plane, we spatially partition the system into two subsystems defined as: $x>0$ is Subsystem A; $x<0$ is Subsystem B. The y-direction obeys periodic boundary conditions while the boundary condition of x-direction is unimportant. Suppose that the lattice constants in x-direction and y-direction are $a$ and $b$, respectively. Therefore, $C$ must be function of momentum $k_y\in (-\pi/b,\pi/b)$. Hence from now on, all site indices $i,j,\cdots$ denote merely the x-coordinates, and $C=\{C^{k_y}_{i,a;j,b}\}$. Schematically, $C$ is decomposed into four parts:
\begin{eqnarray}
C=\left(
  \begin{array}{cc}
    C_{AA} & C_{AB} \\
    C_{BA} & C_{BB} \\
  \end{array}
\right)
\end{eqnarray}
where, $C_{AA}=C^\dagger_{AA}$, $C_{BB}=C^\dagger_{BB}$, $C_{AB}=C^\dagger_{BA}$. Each part is labeled by a given momentum $k_y$. The requirement $C^2=C$ leads to£º
\begin{align}
C_{AA} (1-C_{AA})&=C_{AB}C^\dagger_{AB} \label{eq-identity1} \\
C_{BB}(1-C_{BB})&=C^\dagger_{AB}C_{AB}\label{eq-identity2}\\
C_{AB}(1-C_{BB})&=C_{AA}C_{AB}\label{eq-identity3}
\end{align}
Suppose $C_{AA}$ is $N_A\times N_A$ and $C_{BB}$ is $N_B\times N_B$. Then, $C_{AB}$ is $N_A\times N_B$ while $C_{BA}$ is $N_B\times N_A$. Let's apply SVD (single-valued decomposition)  on off-diagonal submatrix $C_{AB}$:
\begin{align}
  C_{AB}=\mathcal{U}^\dagger D \mathcal{V}
\end{align}
where, $\mathcal{U}$ and $\mathcal{V}$ are $N_A\times N_A$ and $N_B\times N_B$ unitary matrices. $D$ is a $N_A\times N_B$ matrix with non-negative real diagonal elements and zero else. Accordingly,
\begin{align}
  C_{BA}=C^\dagger_{AB}=\mathcal{V}^\dagger D^T \mathcal{U}
\end{align}
with superscript $T$ denoting ``transpose'' operation. It is thus straightforward to obtain: $C_{AB}C^\dagger_{AB}=\mathcal{U}DD^T\mathcal{U}$ and $C^\dagger_{AB}C_{AB}=\mathcal{V}D^TD\mathcal{V}$. Then, Eqs. (\ref{eq-identity1}, \ref{eq-identity2}, \ref{eq-identity3}) are transformed to:
\begin{align}
  \mathcal{U}C_{AA}(1-C_{AA})\mathcal{U}^\dagger&=DD^T\\
  \mathcal{V}C_{BB}(1-C_{BB})\mathcal{V}^\dagger&=D^TD\\
 D\mathcal{V} (1-C_{BB}) \mathcal{V}^\dagger &= \mathcal{U}C_{AA}\mathcal{U}^\dagger D
\end{align}
We define: $\widetilde{C}_{AA}\equiv\mathcal{U}C_{AA}\mathcal{U}^\dagger$, $\widetilde{C}_{BB}\equiv\mathcal{V}C_{BB}\mathcal{V}^\dagger$. Then,
\begin{align}
  \widetilde{C}_{AA}(1-\widetilde{C}_{AA})&=DD^T\label{eq-identity4}\\
  \widetilde{C}_{BB}(1-\widetilde{C}_{BB})&=D^TD\label{eq-identity5}\\
 D(1-\widetilde{C}_{BB})  &= \widetilde{C}_{AA}  D\label{eq-identity6}
\end{align}
Since both $DD^T$ and $D^TD$ are diagonal, the SVD operation simultaneously diagonalizes the four parts of $C$. That is, $\widetilde{C}_{AA}=diag(\lambda_1,\cdots,\lambda_{N_A})$. Furthermore, without lost of generality, we assume $N_A\leq N_B$. According to Eq. (\ref{eq-identity4}), $DD^T=diag(\lambda_1(1-\lambda_1),\cdots,\lambda_{N_A}(1-\lambda_{N_A}))$. Then, $D^TD=diag(\lambda_1(1-\lambda_1),\cdots,\lambda_{N_A}(1-\lambda_{N_A}),0,\cdots,0)$ where there are $N_B-N_A$ zero diagonal terms in addition. The $N_A\times N_B$ matrix $D$ has $N_A$ diagonal elements $\{\sqrt{\lambda_1(1-\lambda_1)},\cdots,\sqrt{\lambda_{N_A}(1-\lambda_{N_A})}\}$. According to Eq. (\ref{eq-identity6}), the solution to Eq. (\ref{eq-identity5}) is only one case: $\widetilde{C}_{BB}=diag(1-\lambda_1,\cdots,1-\lambda_{N_A},0,\cdots,0)$.  Therefore,
\begin{align}
  C=\left(
    \begin{array}{cc}
      \mathcal{U}^\dagger & 0 \\
     0 & \mathcal{V}^\dagger \\
    \end{array}
  \right)\left(
    \begin{array}{cc}
      \widetilde{C}_{AA} &       \widetilde{C}_{AB} \\
           \widetilde{C}_{BA} &       \widetilde{C}_{BB}\\
    \end{array}
  \right)\left(
    \begin{array}{cc}
      \mathcal{U} & 0 \\
     0 & \mathcal{V} \\
    \end{array}
  \right)
\end{align}
Each column of $\left(
    \begin{array}{cc}
      \mathcal{U} & 0 \\
     0 & \mathcal{V} \\
    \end{array}
\right)$ forms an orthonormal vector in $N_A+N_B$ dimensional vector space. Since the off-diagonal submatrices are zero, $\mathcal{U}$ and $\mathcal{V}$ independently form two sets of orthonomal vectors. These two sets are again orthogonal to each other:
\begin{align}
  \mathcal{U}&=\left\{|n_A\rangle\right\},\,\,n=1,2,\cdots,N_A\\
  \mathcal{V}&=\left\{|n_B\rangle\right\},\,\,n=1,2,\cdots,N_B\\
  &\langle n_A|n'_B\rangle=0.
\end{align}
In 2nd-quantized language, each single-particle state $|n_A\rangle$ ($|n_B\rangle$) can be created from the \emph{single-body} vacuum state $|0_{nA}\rangle$ ($|0_{nB}\rangle$)
by a fermionic creation operator $\Gamma^\dagger_{nA}$ ($\Gamma^\dagger_{nB}$):
\begin{align}
 |n\rangle_A= \Gamma^\dagger_{nA}|0_{nA}\rangle\,,\,\,|n\rangle_B=\Gamma^\dagger_{nB}|0_{nB}\rangle
\end{align}
Their corresponding eigenvalues $``\lambda_n$'' and ``$1-\lambda_n$'' are the probability of occupying the states, respectively. These operators satisfy the anticommutation relation:
\begin{align}
  \{\Gamma_{nA},\Gamma^\dagger_{n'A}\}=\delta_{nn'}\,\,,\,\,  \{\Gamma_{nB},\Gamma^\dagger_{n'B}\}=\delta_{nn'}\,\,,
\end{align}
and zero for else.

Let's define the many-body vacuum states:
\begin{align}
  |0_A\rangle\equiv \bigotimes_{n}|0_{nA}\rangle\,\,,\,\,  |0_B\rangle\equiv \bigotimes_{n}|0_{nB}\rangle\,\,,\,\,
\end{align}
Let's also define the many-body \emph{cut-Groundstates} of subsystem A and B, respectively:
\begin{align}
  |\Omega_A\rangle\equiv\left(\prod_{n\geq\frac{1}{2}}\Gamma^\dagger_{nA}\right)|0_A\rangle\,\,,\,\, \,\,\, |\Omega_B\rangle\equiv\left(\prod_{\lambda_n<\frac{1}{2}}\Gamma^\dagger_{nB}\right)|0_B\rangle\,,\label{eq-cut}
\end{align}
where the ordering of fermionic operators is presumed to be ``$\Gamma_{N_A}\cdots\Gamma_2\Gamma_1$''. From the previous page, the ground state $|\text{GS}\rangle$ of the whole system is written as:
\begin{align}
  |\text{GS}\rangle=\left[\prod_{n}\left( \sqrt{\lambda_n}\Gamma^\dagger_{nA}+\sqrt{1-\lambda_n}\Gamma^\dagger_{nB} \right)\right]|0_A\rangle\otimes|0_B\rangle
\end{align}
which can be re-expressed as:
\begin{widetext}
\begin{align}
  |\text{GS}\rangle&=\prod_{n}\left( \sqrt{\lambda_n}\Gamma^\dagger_{nA}+\sqrt{1-\lambda_n}\Gamma^\dagger_{nB} \right)\cdot |0_A\rangle\otimes|0_B\rangle\nonumber\\
  &=\prod_{\lambda_n\geq\frac{1}{2}}\left( \sqrt{\lambda_n}\Gamma^\dagger_{nA}+\sqrt{1-\lambda_n}\Gamma^\dagger_{nB} \right)\,\cdot\,\prod_{\lambda_n<\frac{1}{2}}\left( \sqrt{\lambda_n}\Gamma^\dagger_{nA}+\sqrt{1-\lambda_n}\Gamma^\dagger_{nB} \right)\cdot |0_A\rangle\otimes|0_B\rangle\nonumber\\
  &\propto \prod_{\lambda_n\geq\frac{1}{2}}\left( \Gamma^\dagger_{nA}+\frac{\sqrt{1-\lambda_n}}{\sqrt{\lambda_n}}\Gamma^\dagger_{nB} \right)\,\cdot\,\prod_{\lambda_n<\frac{1}{2}}\left( \Gamma^\dagger_{nB}+\frac{\sqrt{\lambda_n}}{\sqrt{1-\lambda_n}}\Gamma^\dagger_{nA} \right)\cdot |0_A\rangle\otimes|0_B\rangle\nonumber\\
  &=\prod_{\lambda_n\geq\frac{1}{2}}\left( \Gamma^\dagger_{nA}+\frac{\sqrt{1-\lambda_n}}{\sqrt{\lambda_n}}\Gamma^\dagger_{nB} \Gamma_{nA}\Gamma^\dagger_{nA} \right)\,\cdot\,\prod_{\lambda_n<\frac{1}{2}}\left( \Gamma^\dagger_{nB}+\frac{\sqrt{\lambda_n}}{\sqrt{1-\lambda_n}}\Gamma^\dagger_{nA}  \Gamma_{nB}\Gamma^\dagger_{nB} \right)\cdot |0_A\rangle\otimes|0_B\rangle\label{line1}\\
  &=\prod_{\lambda_n\geq\frac{1}{2}} \left[\left(1+\frac{\sqrt{1-\lambda_n}}{\sqrt{\lambda_n}}\Gamma^\dagger_{nB} \Gamma_{nA}\right)\Gamma^\dagger_{nA} \right]\cdot\,\prod_{\lambda_n<\frac{1}{2}}\left[\left(1+\frac{\sqrt{\lambda_n}}{\sqrt{1-\lambda_n}}\Gamma^\dagger_{nA} \Gamma_{nB}\right)\Gamma^\dagger_{nB}\right]\cdot |0_A\rangle\otimes|0_B\rangle\nonumber\\
  &=e^{\sum_{\lambda_n\geq\frac{1}{2}}\frac{\sqrt{1-\lambda_n}}{\sqrt{\lambda_n}}\Gamma^\dagger_{nB} \Gamma_{nA}}\left(\prod_{\lambda_n\geq\frac{1}{2}} \Gamma^\dagger_{nA} \right)\cdot\,e^{\sum_{\lambda_n<\frac{1}{2}}\frac{\sqrt{\lambda_n}}{\sqrt{1-\lambda_n}}\Gamma^\dagger_{nA} \Gamma_{nB}}\left(\prod_{\lambda_n<\frac{1}{2}} \Gamma^\dagger_{nB} \right)\cdot |0_A\rangle\otimes|0_B\rangle\nonumber\\
  &=e^{\sum_{\lambda_n\geq\frac{1}{2}}\frac{\sqrt{1-\lambda_n}}{\sqrt{\lambda_n}}\Gamma^\dagger_{nB} \Gamma_{nA}}\cdot\,e^{\sum_{\lambda_n<\frac{1}{2}}\frac{\sqrt{\lambda_n}}{\sqrt{1-\lambda_n}}\Gamma^\dagger_{nA} \Gamma_{nB}}\left(\prod_{\lambda_n\geq\frac{1}{2}} \Gamma^\dagger_{nA} \right)\left(\prod_{\lambda_n<\frac{1}{2}} \Gamma^\dagger_{nB} \right)\cdot |0_A\rangle\otimes|0_B\rangle\label{line2}\\
  &=e^{\sum_{\lambda_n\geq\frac{1}{2}}\frac{\sqrt{1-\lambda_n}}{\sqrt{\lambda_n}}\Gamma^\dagger_{nB} \Gamma_{nA}+\sum_{\lambda_n<\frac{1}{2}}\frac{\sqrt{\lambda_n}}{\sqrt{1-\lambda_n}}\Gamma^\dagger_{nA} \Gamma_{nB}}|\Omega_A\rangle\otimes|\Omega_B\rangle\,.
\end{align}
In Line (\ref{line1}), we have inserted $\Gamma_{nA}\Gamma^\dagger_{nA}|0_A\rangle=(1-\Gamma_{nA}^\dagger\Gamma_{nA})|0_A\rangle=(1-0)|0_A\rangle=|0_A\rangle$, and $\Gamma_{nB}\Gamma^\dagger_{nB}|0_B\rangle=(1-\Gamma_{nB}^\dagger\Gamma_{nB})|0_B\rangle=(1-0)|0_B\rangle=|0_B\rangle$. In Line (\ref{line2}), the operator $\left(\prod_{\lambda_n\geq\frac{1}{2}} \Gamma^\dagger_{nA} \right)$ commutes with $e^{\sum_{\lambda_n<\frac{1}{2}}\frac{\sqrt{\lambda_n}}{\sqrt{1-\lambda_n}}\Gamma^\dagger_{nA} \Gamma_{nB}}$. In last line, the operator $\sum_{\lambda_n\geq\frac{1}{2}}\frac{\sqrt{1-\lambda_n}}{\sqrt{\lambda_n}}\Gamma^\dagger_{nB} \Gamma_{nA}$ commutes with $\sum_{\lambda_n<\frac{1}{2}}\frac{\sqrt{\lambda_n}}{\sqrt{1-\lambda_n}}\Gamma^\dagger_{nA} \Gamma_{nB}$.

Since we are on a cylinder, $k_y$ is good quantum number. Multiplying the above results for each $k_y$, we obtain the true ground state of whole system:
\begin{align}
  |\text{GS}\rangle=\bigotimes_{k_y}|\text{GS}_{k_y}\rangle
\end{align}
where $|\text{GS}_{k_y}\rangle$ is given by adding label $k_y$:
\begin{align}
  \Gamma_{nA}\rightarrow \Gamma_{nA;k_y}\,\,,\,\,\,  \Gamma_{nB}\rightarrow \Gamma_{nB;k_y}\,\,,\,\,\,  |\text{GS}\rangle\rightarrow  |\text{GS}_{k_y}\rangle\,\,,\,\,\,  \lambda_{n}\rightarrow  \lambda_{n}^{k_y}\,\,,\,\,\,
\end{align}
so that,
\begin{align}
  |\text{GS}_{k_y}\rangle&=\prod_{n}\left( \sqrt{\lambda^{k_y}_n}\Gamma^\dagger_{nA;k_y}+\sqrt{1-\lambda^{k_y}_n}\Gamma^\dagger_{nB;k_y} \right)\cdot |0_A\rangle\otimes|0_B\rangle\nonumber\\
  &=\exp\left\{\sum_{\lambda^{k_y}_n\geq\frac{1}{2}}\left({\lambda^{k_y}_n}^{-1}-1\right)^{\frac{1}{2}}\Gamma^\dagger_{nB;k_y} \Gamma_{nA;k_y}+\sum_{\lambda^{k_y}_n<\frac{1}{2}}\left({\lambda^{k_y}_n}^{-1}-1\right)^{-\frac{1}{2}}\Gamma^\dagger_{nA;k_y} \Gamma_{nB;k_y}\right\}|\Omega_{A,k_y}\rangle\otimes|\Omega_{B,k_y}\rangle\,.
\end{align}
Alternatively, we can expand the exponential in Taylor series by considering anticommutation algebra:
\begin{align}
  |\text{GS}_{k_y}\rangle=&|\Omega_{A,k_y}\rangle\otimes|\Omega_{B,k_y}\rangle+\left\{\sum_{\lambda^{k_y}_n\geq\frac{1}{2}}\left({\lambda^{k_y}_n}^{-1}-1\right)^{\frac{1}{2}}\Gamma^\dagger_{nB} \Gamma_{nA;k_y}+\sum_{\lambda^{k_y}_n<\frac{1}{2}}\left({\lambda^{k_y}_n}^{-1}-1\right)^{-\frac{1}{2}}\Gamma^\dagger_{nA;k_y} \Gamma_{nB;k_y}\right\}|\Omega_{A,k_y}\rangle\otimes|\Omega_{B,k_y}\rangle\nonumber\\
  =&|\Omega_{A,k_y}\rangle\otimes|\Omega_{B,k_y}\rangle\nonumber\\
  &+\sum_{\lambda^{k_y}_n<\frac{1}{2}}\left({\lambda^{k_y}_n}^{-1}-1\right)^{-\frac{1}{2}} \Gamma^\dagger_{nA;k_y} |\Omega_{A,k_y}\rangle\otimes\Gamma_{nB;k_y}|\Omega_{B,k_y}\rangle+\sum_{\lambda^{k_y}_n\geq\frac{1}{2}}\left({\lambda^{k_y}_n}^{-1}-1\right)^{\frac{1}{2}} \Gamma_{nA;k_y} |\Omega_{A,k_y}\rangle\otimes\Gamma^\dagger_{nB;k_y}|\Omega_{B,k_y}\rangle\label{eq-schmidt}
\end{align}
By definition, the following normalization is satisfied:
\begin{align}
\langle\Omega_{A,k_y}|\Omega_{A,k_y}\rangle=1\,,\,\langle\Omega_{B,k_y}|\Omega_{B,k_y}\rangle=1\,.
\end{align}
such that, for $\lambda^{k_y}_{n}<\frac{1}{2}$,
\begin{align}
\langle\Omega_{A,k_y}|\Gamma_{nA;k_y}\Gamma^\dagger_{nA;k_y} |\Omega_{A,k_y}\rangle&=\langle\Omega_{A,k_y}|\left(1-\Gamma^\dagger_{nA;k_y}\Gamma_{nA;k_y}\right) |\Omega_{A,k_y}\rangle=\langle\Omega_{A,k_y}|\left(1-0\right) |\Omega_{A,k_y}\rangle=1\,,\nonumber\\
\langle\Omega_{B,k_y}|\Gamma^\dagger_{nB;k_y}\Gamma_{nB;k_y}|\Omega_{B,k_y}\rangle&=\langle\Omega_{B,k_y}|1|\Omega_{B,k_y}\rangle=1\nonumber
\end{align}
and, for $\lambda^{k_y}_{n}\geq\frac{1}{2}$,
\begin{align}
\langle\Omega_{B,k_y}|\Gamma_{nB;k_y}\Gamma^\dagger_{nB;k_y} |\Omega_{B,k_y}\rangle&=\langle\Omega_{B,k_y}|\left(1-\Gamma^\dagger_{nB;k_y}\Gamma_{nB;k_y}\right) |\Omega_{B,k_y}\rangle=\langle\Omega_{B,k_y}|\left(1-0\right) |\Omega_{B,k_y}\rangle=1\,,\nonumber\\
\langle\Omega_{A,k_y}|\Gamma^\dagger_{nA;k_y}\Gamma_{nA;k_y}|\Omega_{A,k_y}\rangle&=\langle\Omega_{A,k_y}|1|\Omega_{A,k_y}\rangle=1\nonumber
\end{align}
\end{widetext}

Therefore, we can read out the ``\emph{entanglement energy} $\varepsilon_{n}(k_y)$'' (which forms \emph{entanglement spectrum}) directly from Eq. (\ref{eq-schmidt}):
\begin{align}
  e^{- {\varepsilon_n(k_y)}}&=\left({\lambda^{k_y}_n}^{-1}-1\right)^{-1}  \,.
  \end{align}
 
which agrees with Eq. (\ref{peschel}).

\section{Exponential decay rate of the Wannier functions }
\label{fourieranalytic}

In this appendix, we shall show prove that the Fourier coefficients $\psi(x)$ in Eq. \ref{fourierresult}, i.e. 

\begin{equation}
\psi(x)\propto \int dk e^{ikx}e^{i\theta(k)}\psi(k) 
\end{equation}

decay like $\psi(x)\sim e^{-gx}$, where $e^{i\theta(k)}\psi(k)$ has a singularity at $Im(k)=g$, but is analytic for $Im(k)<g$. This is an important theorem that our key result Eq. \ref{key} prominently relies on. A similar proof can already be found in for instance Refs. \onlinecite{leeandy2014} or \onlinecite{kohn1959}, though in different contexts. Here, we shall reproduce it in a way tailored to our context.

Since $\psi(k)$ is an eigenfunction of the hamiltonian $h(k)$ (up to a phase factor), it belongs to a degenerate eigenspace when there the gap closes. Consider the analytic continuation (with abuse of notation) of $\psi(k)$ into $\psi(z) =\psi(e^{ik})$:
\begin{equation}  \psi(z)= \sum_{x\ge 0} \frac{\psi(x)}{2} \left(z^x + \frac{1}{z^x}\right) \label{fourierseries}\end{equation}

Due to the Theorem of Monera (check) and the fact that $h(z)$ is real on the unit circle $|z|=1$, $\psi(z)$ necessarily has a singularity (pole or branch point) inside the unit circle. Let $z_0$ be the singularity of largest magnitude inside the unit circle. We want to show that 

\begin{equation} |\psi(x)|\sim |z_0|^x =e^{-gx}\label{decaykproof}\end{equation}
up to a proportionality factor, where $|z_0| <1$.  In particular, there is a constant $C$ such that $|\psi_x|<C|z_0|^x$. This is a known result\cite{kohn1959,he2001}, and in the next paragraph we sketch a simple derivation suitable for our context. 

Since $\psi(z)$ is analytic for $|z|>|z_0|$ within the unit circle, the series Eq. \ref{fourierseries} must converge in that region.   As Eq. (\ref{decaykproof}) must hold for some value of $|z_0|$ for this series to converge at all inside the unit circle, 
for $z$ such that $|z_0| <|z|<1$, \begin{equation}
|\psi(z)| < \sum_{x\ge 0} \frac{|\psi(x)|}{|z|^x} < C \sum_{x\ge 0} \left|\frac{z_0}{z}\right|^x < \infty
\end{equation}In addition, $\psi(z)$ fails to be analytic at $z_0$, so the above series must diverge when $|z|=|z_0|$.  
This implies that $|\psi(x)|$ must asymptotically decay like $|z_0|^x$, rhus proving Eq. (\ref{decaykproof}).

\section{Details on the Entanglement Spectrum of the Dirac model} 
\label{app:dirac}
There are two main mathematical quantities to determine in Eq. \ref{key}: The monotonically increasing function $f(g(k_y))$, and the Wannier polarization $X_\pm (k_y)$, where $\pm$ label the ES corresponding to the two entanglement cuts.

\subsection{Determination of linear ansatz parameters}

To a first approximation, the function $f(g)$ is given by
\begin{equation}
f(g)= (2+A)g+J
\label{linearansatzapp}
\end{equation}
where $A$ and $J$ are parameters. In the case of the Dirac models, there indeed exists two points $k_y=0$ and $\pi$ where the entanglement spectra can be rigorously solved. $A$ and $J$ can thus be obtained by fitting Eqs. \ref{key} and \ref{linearansatzapp} with the exact results.

At $k_y=0$ or $\pi$, the hamiltonian is given by $H=\sigma\cdot d$, where $d(k)=(m+\cos k_x+\cos k_y,\sin k_x,\sin k_y)=(m\pm 1 + \cos k_x,\sin k_x,0)$, i.e. only $d_1$ and $d_2$ are nonzero. In such cases, there exists exact analytic results for the asymptotic spacing between entanglement eigenvalues  
\begin{eqnarray}
&&\lim_{n\rightarrow \infty}(\epsilon_{n+1}-\epsilon_n)\notag\\
&=&(3-sgn(-m\mp 1-2))\frac{\pi}{2}\frac{I(\sqrt{1-|(m\pm 1 )/2|^2})}{I(|(m\pm 1 )/2|)}\notag\\
\label{toeplitzspacing0}
\end{eqnarray}
where the $\mp$ refers to $k_y=0$ or $k_y=\pi$. This impressive result from Eq. \ref{noncritical_spacing} will be explained in more detail later in this Appendix; here we just mention that $A$ and $J$ can be by comparing it with $\epsilon_{n+1}-\epsilon_n\approx (2+A)g(k_y)+J$, where $k_y=0$ or $\pi$. 

To find $g(0)$ and $g(\pi)$, we solve for $h(k_0,k_y)=d_1^2+d_2^2=0$, and identify $g$ with $Im(k_0)$. It is easily shown that the gap closes at complex $k_0=(1+sgn(P))\pi/2 + i\cosh^{-1}|P|$, where 
\begin{equation} P=\frac{2+m^2+2m\cos k_y}{(\cos k_y+m)}=\frac{2+m^2\pm 2m}{m\pm 1}\end{equation} 
Hence $g(0)=\frac{1}{m+1}+m+1$ and $g(\pi)=\frac{1}{m-1}+m-1$, and $A,J$ can be easily obtained. 





\subsubsection{The exact ES for certain 2-band hamiltonians through Toeplitz Matrices}
\label{2bandtoeplitz}

Here we discuss some known exact results for the eigenspectrum of 2-band models, with the goal of obtaining Eq. \ref{toeplitzspacing0}. Consider $d_1(k_x),d_2(k_x)$ (with $k_y$ as a parameter) of the form

\begin{equation} d_1(k_x) = \cos k_x - \alpha/2 \end{equation} 
\begin{equation} d_2(k_x) = \gamma \sin k_x \end{equation}

with $\gamma\neq 0$ and $\alpha>0$ so the system is gapped. For our Dirac model at $k_y=0$ or $\pi$, $\alpha=-2m\mp 2$ and $\gamma=1$. Although the exact eigenspectrum of \emph{block} Toeplitz matrices are notoriously hard to compute, in the current case a brilliant solution was found by \onlinecite{its2008}. The asymptotic (large $L_A$) spacing between the eigenvalues were found to be
\begin{equation}
\lim_{n\rightarrow \infty}(\epsilon_{n+1}-\epsilon_n)=(3-sgn(\alpha-2))\frac{\pi}{2}\frac{I(\sqrt{1-\kappa^2})}{I(\kappa)} 
\label{noncritical_spacing}
\end{equation}

which tends towards a constant, unlike those of critical 1-D systems which goes like $\propto \frac{1}{\log L_A}$. Here $I(\kappa)=\int_0^1 \frac{dx}{\sqrt{(1-x^2)(1-\kappa^2x^2)}}$ is the complete elliptic integral of the first kind\cite{jeffreys1999book,stone2009book}, and 
\begin{itemize}
\item $\kappa=\sqrt{\alpha^2/4+\gamma^2-1}/\gamma$ if $4(1-\gamma^2)<\alpha^2<4$
\item $\kappa=\sqrt{(1-\alpha^2/4-\gamma^2)/(1-\alpha^2/4)}$ if $\alpha^2<4(1-\gamma^2)$
\item $\kappa=\gamma/\sqrt{\alpha^2/4+\gamma^2-1}$ if $\alpha>2$
\end{itemize}
with $\kappa'=\sqrt{1-\kappa^2}$. For the Dirac model, the first (third) case applies when $(m\pm 1)^2<1$ ($(m\pm 1)^2>1$).  
As a bonus, we also have exact expression for the entanglement entropy
\begin{equation} S_A=\frac{1}{6}\left(log\frac{\kappa^2}{16\kappa'}+\left(1-\frac{\kappa^2}{2}\right)\frac{4I(\kappa)I(\kappa')}{\pi}\right)+log2 \end{equation}
for $\alpha<2$, and 
\begin{equation} S_A=\frac{1}{12}\left(log\frac{16}{\kappa^2\kappa'^2}+\left(\kappa^2-\kappa'^2\right)\frac{4I(\kappa)I(\kappa')}{\pi}\right) \end{equation}
for $\alpha>2$. All these results can be obtained via a detailed analysis of the pole positions of $\phi(z)$ in Eq. \ref{toeplitzpoly}. 
Note that $S_A$ tends to a constant asymptotically, unlike in the critical case. The Entanglement Entropy of the whole system will then by proportional to the length of the cut $L_y$, in agreement with well-known area laws\cite{wolf2006,verstraete2006,plenio2005,li2008,swingle2010}.


\subsection{Wannier Polarization for the Entanglement Spectrum}

For a single occupied band, the Wannier polarization is given by\cite{qi2011}
\begin{equation}  X(k_y)=\frac{1}{2\pi}\int^{2\pi}_0 \psi(k_x,k_y)^\dagger\partial_{k_x}\psi(k_x,k_y) dk_x  \end{equation} 
where $\psi(k_x,k_y)$ is the occupied (lower energy) eigenstate. For the Dirac model, we explicitly have
\begin{equation} \psi(k_x,k_y)=\frac{1}{N}(-sink_x + isink_y, m(+cosk_x+cosk_y)+\lambda)^T\end{equation}  
Here the normalization factor $N=\sqrt{2\lambda((m+cosk_x+cosk_y)+\lambda)}$ with $\lambda=\sqrt{sin^2k_x+sin^2k_y+(m+cosk_x+cosk_y)^2}$. A few simplifications are in order. We write $\psi=(a+bi)/N$, where $a$ and $b$ are real vectors. As $|\psi|^2=a^2+b^2=1$ is a constant, the real parts of $\psi^\dagger\partial_{k_x}\psi$ must disappear. Hence 
\begin{eqnarray}
\psi^\dagger\partial_{k_x}\psi&=&i ((b/N)\partial_{k_x}(a/N)-(a/N)\partial_{k_x}(b/N))\notag \\
&=&\frac{i}{N^2}(b\partial_{k_x}a-a\partial_{k_x}b)\notag\\
&=&\frac{i}{N^2}(-sink_y\partial_{k_x}sink_x)\notag\\
&=&\frac{-i sink_ycosk_x}{2\lambda((m+cosk_x+cosk_y)+\lambda)}
\label{derivation1}
\end{eqnarray}
Thus the exact integral expression for the polarization is 
\begin{widetext}
\begin{equation}X(k_y)= \int^{2\pi}_0\frac{-sink_ycosk_x}{2\pi\sqrt{sink_x^2+sink_y^2+ (m+cosk_x+cosk_y)^2}(\sqrt{sink_x^2+sink_y^2+ (m+cosk_x+cosk_y)^2}+m+cosk_x+cosk_y)}dk_x
\label{exactqah}
\end{equation}
\end{widetext}
This is a complicated but tractable integral, and its full form must be retained to maintain accuracy over all values of $m$, especially in the topologically nontrivial regime $|m|<2$ where the polarization has a winding of $\pm 1$ upon $k_y\rightarrow k_y+2\pi$. In the ES given by Eq. \ref{key}, we use $X(k_y)$ and $2\pi-X(k_y)$ for the spectra corresponding to the two different edges.

\section{Primer on Toeplitz Matrices}
\label{app:history}

Here, we provide an overview of the history and development of Toeplitz Matrices, so as to put our asymptotic estimates of the spectra of Toeplitz eigenvalues in better perspective. We have included it as a separate appendix so as not to distract readers from the goal of this work, which is to quantitatively understand the asymptotic properties of entanglement spectra.

Toeplitz matrices are finite-sized matrices that have translational symmetry along each diagonal.  They appear in a wide variety of applications, from the thermodynamic limit of the 2D classical Ising model and its generalizations\cite{ising1,ising2,ising3}, various spin chain models\cite{its2008, its2009,keating2005}, dimer models\cite{basordimer2007}, impenetrable bose gas systems\cite{ffapps} to full counting statistics  and certain non-equilibrium phenomena\cite{ivanov2010phase,noneq}. In a celebrated result by Potts and Ward\cite{ising4}, the spin-spin correlator of the 2D Ising model is expressed as a Toeplitz determinant. In other settings, the asymptotic limits of the eigenvalues of Toeplitz matrices are essential in the calculation of the entanglement spectrum and entropy, such as the XX and XY quantum spin chains and their equivalent free-fermion problems\cite{its2009}. Of more exigent physical importance is the use of Toeplitz determinants in computing the correlation functions of dimer models that arise in high-temperature superconductors\cite{basordimer2007}. Such models, which are equivalent to certain 2D Ising models\cite{kasteleyn1961statistics,kasteleyn1963dimer,stephenson1964ising, fisher1966dimer}, have been used to study the possibility of realizing Anderson's RVB liquid in valence-bond dominated phases\cite{fendley2002classical,moessner2003,rokhsar1988}.  More recently, Toeplitz matrices have also been studied in the context of quantum noise, for instance through the calculation of the full counting statistics of 1-D fermions\cite{abanov2011quantum} or their non-equilibrium interactions via bosonization in the framework of the Keldysh action formalism\cite{noneq}. 
 
In these abovementioned applications, quantities of physical interest are usually computed in the thermodynamic limit, where the size of the finite Toeplitz matrices tend to infinity. In this limit, however, the finite Toeplitz matrices do \emph{not} converge to truly infinite Toeplitz matrices whose spectra can be trivially obtained. Intuitively, this is because finite Toeplitz matrices, no matter how large, will always contain "edges" that nontrivially modify the original spectrum and eigenvectors. This fact is prominently illustrated in the exemplary case of topological insulators, where the Toeplitz matrix is taken to be the real-space hamiltonian. When the Toeplitz matrix is made finite by imposing open boundary conditions, the nontrivial edge eigenstates that appear have distinct energy dispersions from those bulk eigenstates belonging to the original infinite Toeplitz matrix.

As such, a lot has been studied about the asymptotic properties of Toeplitz matrices. The Szeg\"{o} limit theorem\cite{szego} which dates back to 1915 first related the the asymptotics of the determinant of a Toeplitz matrix $T_{ij}=T_{i-j}$ to its symbol $g(k)=\frac{1}{2\pi}\int^{\pi}_{-\pi} T_x e^{ikx}dx$, a quantity that has been introduced in more detail in Section \ref{sec:toeplitz}. Physically, the symbol represents the fourier-space operator corresponding to the Toeplitz matrix representing the truncated real-space version of the same operator. Subsequently, this fundamental 1915 result was extended to the so-called Strong Szeg\"{o} limit theorem requiring much less restrictive assumptions by Kac, Baxter, Hirschman and others\cite{kac,baxter,hirschman}. This result, however, still required the symbol to be continuous with zero winding number. These constraints were relaxed by a series of breakthroughs that follow, thereby opening up the important class of Toeplitz matrices with singular symbols to physical applications\cite{fh2,widom1973,widom1974,widom1975,basor1978,szego2}. Such Toeplitz matrices can physically represent, for instance, flattened hamiltonians acting as projectors to eigensubspaces. In fact, most of the previously mentioned physical applications rely heavily on a class of singular Toeplitz matrices of the Fisher-Hartwig type.

However, relatively little is known about the asymptotic eigenvalue distribution of general Block Toeplitz matrices, i.e. those with matrix-valued symbols $g^{ab}(k)$. They are generalizations of the abovementioned Toeplitz matrices to admit "internal degrees of freedom" which, not surprisingly, will contain vastly richer structure. For instance, Block Toeplitz matrices can represent lattice systems with more than one band, thereby allowing for the possibility of nontrivial topological phenomena\cite{thouless1982}. Exact results for the asymptotic eigenvalue distribution only exists for a special class of $2\times 2$ Block Toeplitz matrices\cite{its2008,basordimer2007}, as already reviewed in Sect. \ref{sec:toeplitz}.. No result on the full asymptotic eigenvalue distribution of general $N \times N$ Block Toeplitz matrices exists to our knowledge, although there has been asymptotic results on the arithmetic mean of their eigenvalues\cite{gutierrez2012}.

Despite their ubiquity, finding the asymptotics of generic Toeplitz matrices remain a notoriously difficult task. The the authors' knowledge, rigorous asymptotic results are not known for the spectra of generic Block Toeplitz matrices with distcontinuous fourier transforms along the diagonals, i.e, those with singular symbols. These are exactly the types of Toeplitz matrices appearing in the entanglement hamiltonians of free-fermion systems. 

As such, it is our hope that our asymptotic bounds on the ES derived via Wannier interpolation will provide some helpful hints on the spectral properties of generic Block Toeplitz Matrices, even those not originally appearing in an entanglement calculation.

\bibliography{fci,entanglement,TI}

\end{document}